\documentclass[aps,prl,amsmath,amssymb,twocolumn,superscriptaddress,showpacs,longbibliography]{revtex4-1}

 \usepackage{graphicx, subfigure}
 \usepackage{amsmath}
 \usepackage{amsfonts}
 \usepackage{amssymb}
 \usepackage{pstricks}
 \usepackage{psfrag}
 \usepackage{epsfig}
\usepackage{helvet}
\usepackage{textcomp}
\usepackage[normalem]{ulem}
\usepackage{hyperref}

\begin{document}
\title{Contactless indentation of a soft boundary by a rigid particle in shear flow}
\author{Alexander Farutin}
\affiliation{Univ. Grenoble Alpes, CNRS, LIPhy, F-38000 Grenoble, France}

\date{\today}

\begin{abstract}
The dynamics of a rigid particle above a fluid-fluid interface in shear flow is studied here numerically and analytically as a function of the downward force applied on the particle.
It is found here that the particle goes below the equilibrium level of the interface for a strong enough downward force.
Such states remain stable under flow, with a fluid film of a well-defined thickness separating the particle from the indented interface.
This result contradicts the classical lubrication theory, which predicts an infinitely large downward force being necessary for the particle to approach the equilibrium level of the interface.
It is found that the classical lubrication approximation is only valid in a narrow range of shear rates, which shrinks to a point when the particle approaches the equilibrium level of the interface.
The gap renormalization model, proposed here, cures this limitation of the classical lubrication theory, showing quantitative agreement with the numerical results when the particle touches the equilibrium level of the interface.
It is found that the gap renormalization model provides a quantitative interpretation of the recent experimental results, including the range of particle heights above the interface for which the classical lubrication approximation breaks down.
\end{abstract}

\maketitle

\paragraph{Introduction}

Lift force is usually discussed in the context of macroscopic objects moving at large Reynolds numbers, such as birds or planes.
This work considers the motion of a rigid particle next to a soft boundary at zero Reynolds number, which is also known to generate lift, as reviewed in\cite{Bureau2023,Rallabandi2024}.
Consequently, a force-free rigid particle near a soft boundary undergoes cross-streamline migration.
Non-inertial cross-streamline migration is also observed for soft particles moving near a rigid boundary\cite{Smart1991,Seifert1999,Cantat1999,Zhao2011,Farutin2013} and plays a crucial role in spatio-temporal organization of suspensions\cite{Wang1998,Hudson2003,Podgorski2011,Narsimhan2013,Grandchamp2013,Zhou2020}, which in turn, affects their rheology\cite{fahraeus1931,Thiebaud2014,Audemar2022}.

The lift of rigid particles near soft boundaries continues to be an important topic of research both from theoretical \cite{Skotheim2004,Beaucourt2004,Skotheim2005,Bertin2022,Jing2024,Jha2024} and experimental\cite{Saintyves2016,Davis2018,Rallabandi2018,Vialar2019,Zhang2020,Zhang2025} points of view.
Lubrication approximation is an widely-used tool in computing the lift force or migration velocity for a rigid particle near a soft wall.
It allows one to calculate the lift force assuming the gap $d$ between the particle and the undeformed boundary to be small compared to the particle size.
Such calculations usually yield an expression proportional to $d^{-\alpha}$\cite{Skotheim2005}, where the exponent $\alpha>0$ strongly depends on the visco-elastic properties of the substrate and on the particle geometry (cylindrical vs. spherical).

A recent experimental measurement of the lift force as a function of $d$ for a rigid sphere translated parallel to a fluid-fluid interface has shown strong deviations from the predictions of the lubrication approximation for small enough $d$\cite{Zhang2025}.
Most notably, the lift force showed saturation as $d$ approached 0, contrary to the divergence predicted by the classical lubrication approximation.

A related problem is the fluid-mediated indentation of a soft substrate by a rigid particle\cite{Kaveh2014,Wang2015,Wang2017}.
In this setting, the particle is subject to a downward force so strong that the particle is pressed into the substrate pushing $d$ into negative values.
At equilibrium, the particle comes in contact with the substrate and indents it according to the applied force and the elastic response of the substrate.
If the particle is moving, the fluid is entrained into the gap between the particle and the substrate so that a fluid film of a well-defined thickness is maintained between the particle and the substrate\cite{Snoeijer2013,Essink2021}.
The thickness of this film can be calculated in the lubrication approximation\cite{Snoeijer2013,Essink2021}.

Studies of lubrication of contacts between soft entities have important practical applications, such as lubrication of mechanical parts\cite{Greenwood2020}, swimming of micro-organisms near a free boundary\cite{Trouilloud2008,Nambiar2022}, the hydrodynamic interaction of blood cells with vessel walls\cite{Davis2018}, and lubrication of articular cartilage in synovial joints\cite{Hou1992}, which support heavy load\cite{Hodge1986} with a surprisingly low friction coefficient\cite{Charnley1960,Jahn2018}.
The measurements of the lift force for a moving rigid probe near a deformable substrate were suggested as a non-invasive probe of the substrate elasticity\cite{Leroy2011,Leroy2012,Kargar2021}.

The goal of this study is to bridge the gap between the limiting cases of positive and negative $d$.
This task is addressed here with high-precision numerical simulations and analytical calculations.
Contrary to the usual approach which focuses on the film between the particle and the substrate\cite{Singh2021,Essink2021}, the computational domain here is chosen to be much larger than the particle size.
With this approach, the lift force can be calculated seamlessly for any sign of $d$ and in the region around $d=0$.
Based on the numerical results, the region of validity of the classical lubrication approximation for $d>0$ is established and the reasons for its breakdown are identified.
A more general analytical model is then proposed and validated by comparison with numerical results.
Finally, the proposed analytical model is used to provide a quantitative interpretation of the recent experimental results.

\paragraph{Model and methods}

\begin{figure}
\begin{center}
\includegraphics[width=\columnwidth]{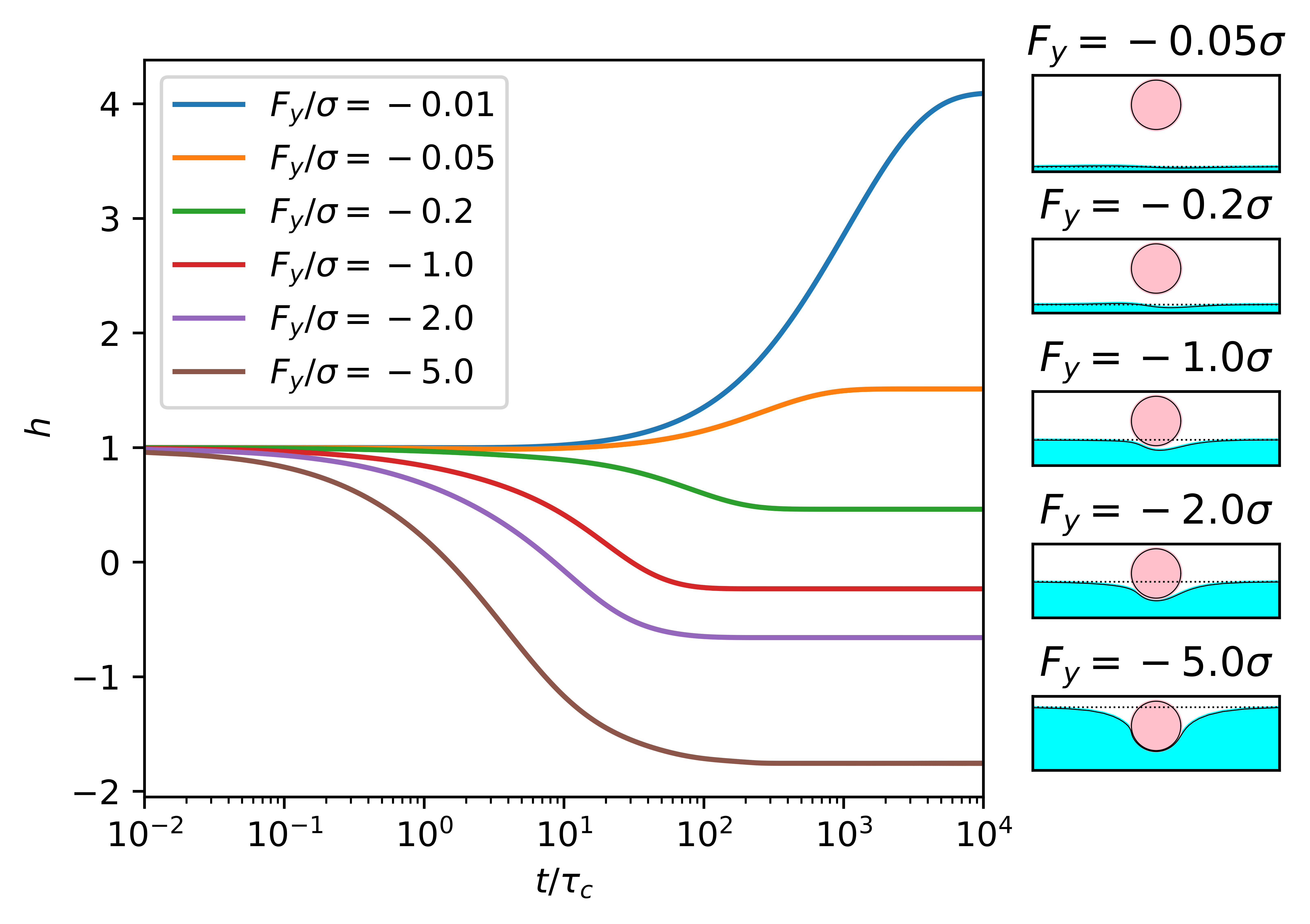}
\caption{\label{fig:time}Time evolution of the particle height above the equilibrium level of the interface. Terminal conformations for several values of the downward force are shown on the right. $B_o=1$, $C_a=0.1$.}
\end{center}
\end{figure}

Here is considered the dynamics of a rigid particle above an interface between two viscous fluids of equal viscosities $\eta$.
The density mismatch between the lower (denser) and upper (lighter) fluids is denoted as $\Delta\rho>0$.
The whole system is subject to shear flow $\boldsymbol u^\infty=\dot\gamma y\boldsymbol e^x$ and the particle experiences an external downward force $\boldsymbol F=F_y\boldsymbol e^y$ ($F_y<0$).
Here $y$ is the axis orthogonal to the interface.
The interface is characterized by the interfacial tension $\sigma$.
The particle is taken as an infinitely long cylinder with its axis aligned with the $z$ direction, which effectively renders the problem two-dimensional.
With this simplification, the particle is represented by its section in the $x$, $y$ plane, a circle of radius $a$, and the shape of the fluid-fluid interface is given by the height function $y_i(x)$, which is independent of $z$.
The coordinates are chosen in such a way that $y_i$ tends to 0 far from the particle.
The particle is taken to be torque-free.
The problem is solved in coordinate system co-moving with the $x$-component of the center $\boldsymbol r_c\equiv (0,y_c)$ of the particle.
The main topic of this study is the relation between the 3 non-dimensional quantities (1) the gap between the particle and the height of the undeformed interface $h(t)\equiv d(t)/a=y_c(t)/a-1$, where $t$ is time, (2) the capillary number $C_a=\dot\gamma\eta a/\sigma$ and (3) the downward force $F_y/\sigma$.
The last non-dimensional parameter is the Bond number $B_o=\Delta\rho g a^2/\sigma$ ($g$ is the free-fall acceleration), which is related to the capillary length $\lambda=aB_o^{-1/2}$.
The time is non-dimensionalized with the capillary relaxation time $\tau_c=\eta a/\sigma$.
Owing to the microscopic size of the particle, the inertial effects are neglected in this study.

The fluid flow is described by Stokes equations, which are solved using boundary integral formulation\cite{Pozrikidis92}, with precision improved by singularity subtraction\cite{Farutin2014b}.
Periodic boundary conditions are imposed along the $x$ direction, with the period $L=1000a$ found to be sufficiently large to approximate the infinite $L$ limit.
Both the interface shape and the boundary forces of the particle are parametrized by Fourier harmonics\cite{Abbasi2022}.
A non-homogeneous parametrization of the interface is used, with about half of the points used to describe the region where the interface and the particle are close to contact.
The forces on the particle boundary are computed directly in the Fourier representation\cite{Abbasi2022}, to avoid having to compute near-singular integrals in the thin-film limit.
The method gives very precise results for fluid film thicknesses above $10^{-3}a$ and even $10^{-4}a$ in some cases, which corresponds to the smallest values used in experiments\cite{Zhang2025}.
More details of the numerical procedure are given in SI\cite{SI}.

\paragraph{Results}

\begin{figure}
    \begin{center}
        \includegraphics[width=\columnwidth]{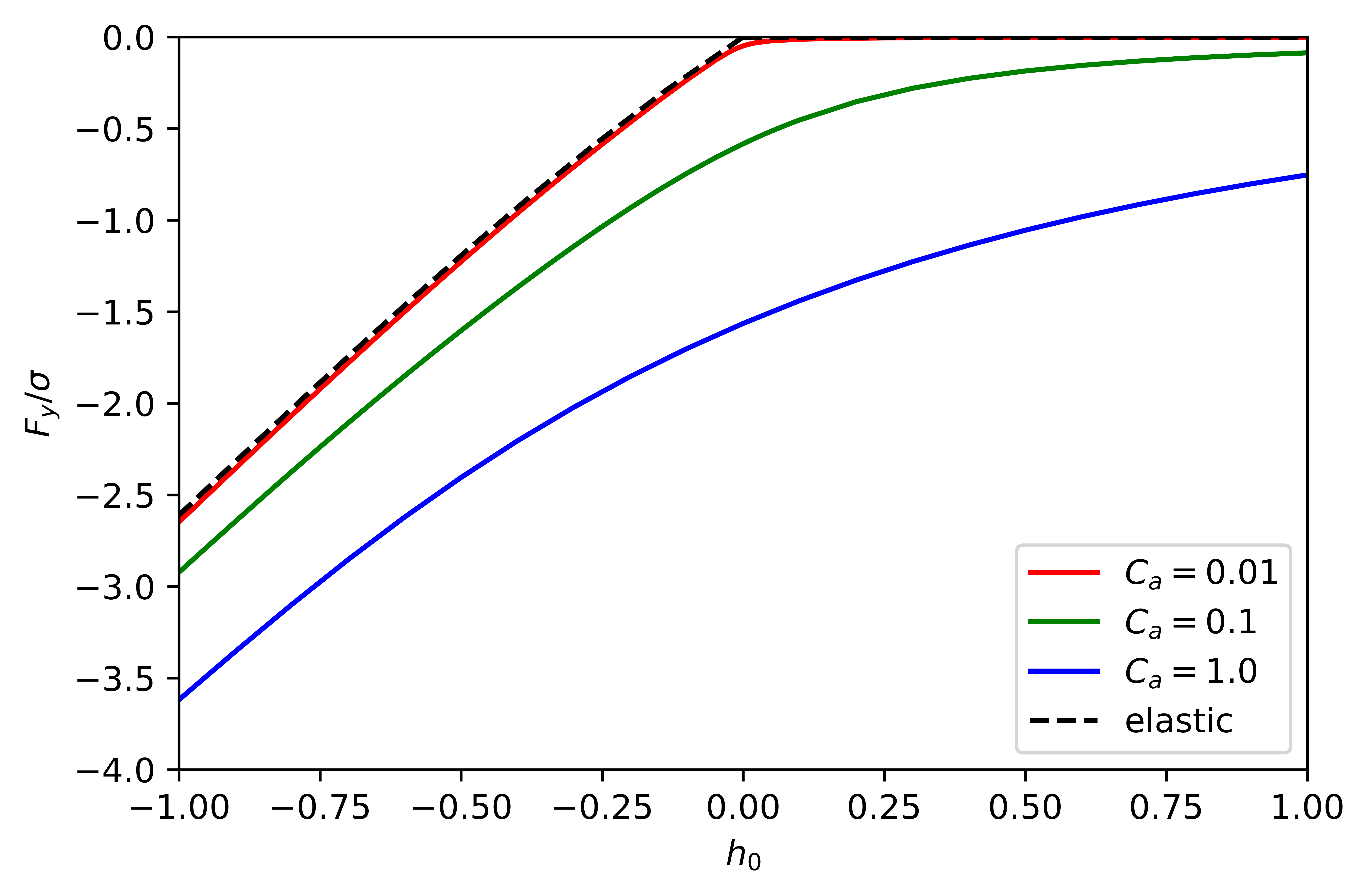}
        \caption{\label{fig:indent} Downward force needed to maintain the particle at a given height relative to the equilibrium level of the interface. $B_o=1$. Dashed lines refer to the purely elastic limit (exact results derived in SI\cite{SI}).
        }
    \end{center}
\end{figure}

Typical simulation results are shown in Fig. \ref{fig:time} and supplementary videos\cite{SI}.
For given $C_a$, $F_y<0$, and initial position of the particle above a flat interface, the particle undergoes vertical migration in the co-moving frame until $h$ reaches a certain value $h_0$, which is a function of $F_y$ and $C_a$.
A remarkable observation is that for a given $C_a$, strong enough downward force leads to negative values of $h_0$.
That is the particle indents the interface in such a way that its lowest part descends below the equilibrium level of the interface.
Nevertheless, the particle never touches the interface and instead slides in the resulting dimple, which moves along with the particle, with a thin film of fluid separating the particle and the interface.
It is important to stress that first, no artificial repulsion between the particle and the interface was used in the numerical procedure and second, the observed dynamics is not a transient effect related to the draining of the fluid from the film between the particle and the interface since it is observed that the particle height and the fluid film thickness reach a stationary value after long enough time.
This means that such states with $h_0<0$ represent a proper dynamical equilibrium in the problem, in which the time-independent thickness of the fluid film is maintained by lubrication pressures.

The systematic analysis of $h_0$ as a function of $F_y$ and $C_a$ is presented in Fig.\ref{fig:indent} ($F_y$ is shown as a function of $h_0$ for several values of $C_a$).
As can be seen, there is no divergence of $F_y$ for $h_0=0$, contrary to the lubrication theory predictions but in agreement with the experimental and numerical results in \cite{Zhang2025}.
Furthermore, the curves for given $C_a$ cross the $h_0=0$ line in a completely regular way, not showing any singularity.
The behavior of $F_y$ as a function of $\dot\gamma$ for a given $h_0$, however, strongly depends on the sign of $h_0$:
For $h_0>0$, the curves approach the line $F_y=0$ as $C_a$ tends to 0, while for $h_0<0$ the curves tend to a well-defined but finite limit.
This limit is nothing but the solution (dashed line) of a purely elastic problem in which the particle indents the interface without fluid, with only contact interaction between the particle and the interface.
Even for flows as weak as $C_a=0.001$, the minimum fluid film thickness remains positive, and shows a linear dependence on $C_a$ for fixed $h_0$ and small $\dot\gamma$.

\paragraph{Discussion}
The numerical results raise several important questions: (1) How does the fluid film between the particle and a strongly indented interface remain stable even under weakest flows? (2) What is the origin of the discrepancy between the classical lubrication theory and the numerical results when $h_0$ tends to 0? (3) Can the short-comings of the classical lubrication approximation be resolved by a more general theoretical framework? These questions are addressed in the following discussion.

\begin{figure}
    \begin{center}
        \includegraphics[width=\columnwidth]{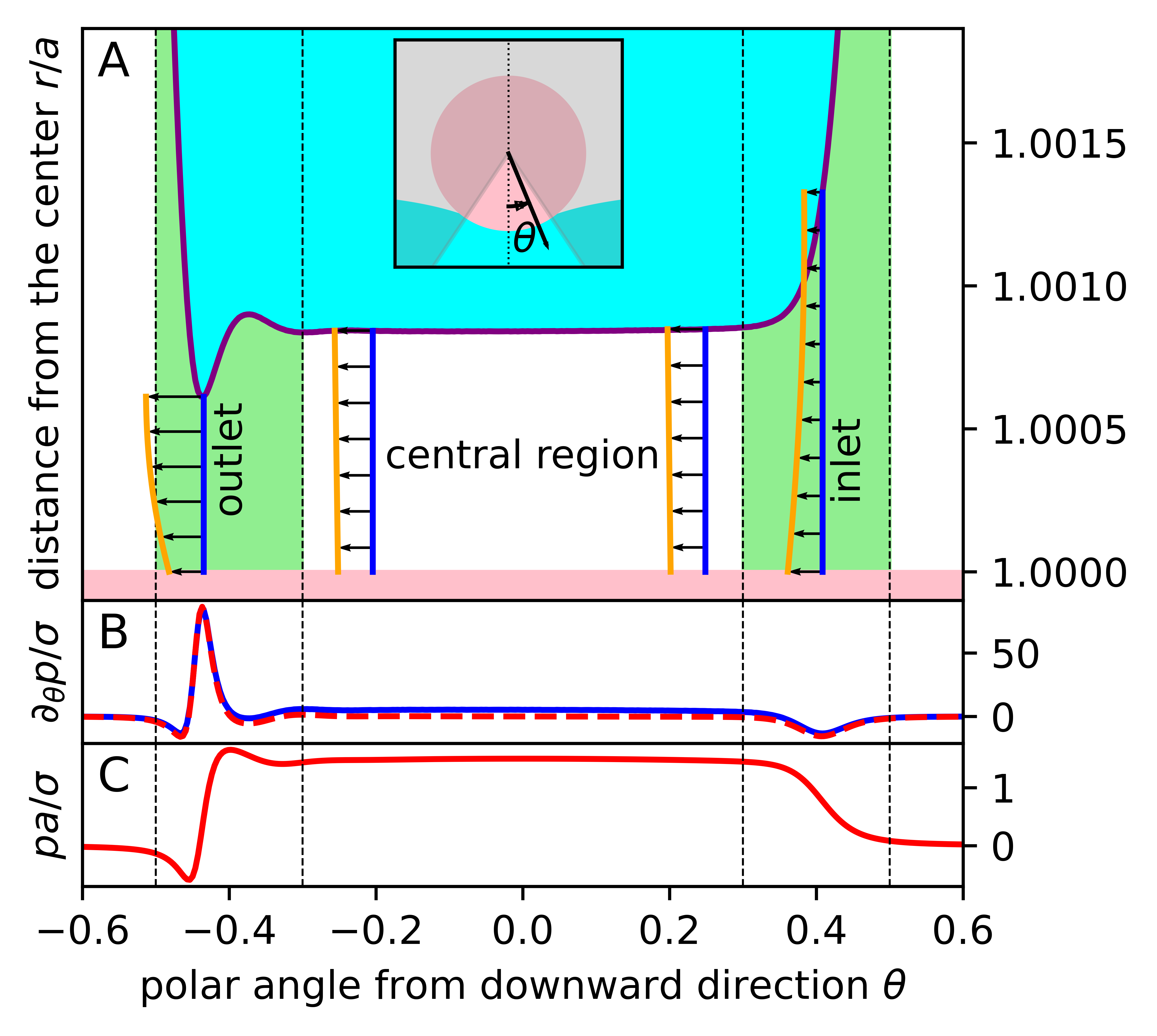}
        \caption{\label{fig:film} Fluid film between the particle and the interface. A: Fluid film geometry in polar coordinates. Black arrows and orange curves show flow profiles (in the reference frame co-moving with the particle) in different sections of the film. Inset marks the region of interest as a non-shaded area. Colors consistent with the main figure. B: Pressure gradient in the film computed from eq. (\ref{Reynolds}) (solid line) and gradient of the interfacial force (dashed line). C: Resulting pressure distribution, computed from the interfacial force. $C_a=0.001$, $B_o=1$. $h_0=-0.5$.
        }
    \end{center}
\end{figure}

The first question is resolved by examining the pressure distribution in the fluid film between the particle and the interface for a given $h_0<0$ in the limit $C_a\rightarrow 0^+$ (Fig.\ref{fig:film}).
In this limit, the flow in the lubrication film and the shape of the interface can be analyzed using the lubrication approximation in polar coordinates.
Taking the $C_a\rightarrow 0^+$ limit, the following equation can be derived (as shown in the SI\cite{SI}) for the fluid film thickness $H$:
\begin{equation}
	\label{Reynolds}
	-\frac{\sigma}{a^3}\partial_{\theta\theta\theta}H=\partial_\theta p/a=\frac{3\eta\Omega(H-H_*)}{H^3},
\end{equation}
where $\theta$ is the polar angle (shown in the inset of Fig.\ref{fig:film}A), $p(\theta)$ is the pressure in the film, $\Omega$ is the angular velocity of the particle, and $H_*$ is a constant related to the total flux in the film.
Equation (\ref{Reynolds}) belongs to the general class of Reynolds-like equations, many of which were analyzed in previous works, including the case of strongly indented substrate.
Similarly to previous works\cite{Essink2021}, it is found here that the fluid film can be decomposed into three regions:
In the central region, $H\simeq H_*$ and according to eq. (\ref{Reynolds}), the fluid pressure shows little variation with $\theta$ (Fig.\ref{fig:film}B).
The pressure in the central region is strong enough to balance the elastic forces of the strongly-deformed interface due to pressure jumps across the inlet and outlet regions (Fig.\ref{fig:film}C).
In the inlet (Fig.\ref{fig:film}), $H$ grows exponentially from $H_*$ as $\theta$ increases, which, according to eq. (\ref{Reynolds}), leads to a pressure jump of order $O(\sigma/a)$ across the inlet region.
In the outlet (Fig.\ref{fig:film}), $H$ shows oscillations of exponentially decreasing amplitude, of which the lowest local minimum makes the strongest contribution to the pressure jump across the inlet region due to the smallness of the $H^3$ term in the denominator.
Since the minimum of $H$ is significantly smaller than $H_*$, the pressure jump across the outlet region is of opposite sign to the pressure jump across the inlet region according to eq. (\ref{Reynolds}).

As derived and validated by numerical results in the SI\cite{SI}, the length of both the inlet and the outlet regions scales as $O(C_a^{1/2})$ and $H_*$ and the minimum film thickness scale as $O(C_a)$, suggesting that the particle and the interface remain separated for arbitrary small $C_a>0$.
The scaling exponents are different from those observed in\cite{Essink2021}, which can be explained by the different visco-elastic properties of the substrate in that work.

The second question is addressed by analyzing the $h_0\rightarrow 0^+$ limit for different values of $C_a$:
For $h_0>0$, $F_y$ is strictly 0 for $C_a=0$ and can also be expanded into a power series of $C_a$:
\begin{equation}
\label{Taylor}
F_y(h_0,C_a)/\sigma=\sum\limits_{k=1}^\infty A_{2k}(h_0)C_a^{2k},
\end{equation}
where only even powers are retained due to the $\dot\gamma\rightarrow-\dot\gamma$ symmetry.
The lubrication approximation provides an expression for $A_2(h_0)$ in the limit of $h_0\rightarrow 0^+$, the leading term of which can be computed exactly in some cases.
The problem at hand is relatively complicated, since the translational and angular velocities of the particle are not known {\it a priori} but need to be related to $\dot\gamma$ from the force and torque balance equations.
These equations include non-negligible contributions from the fluid stresses in the regions that are outside of the thin-film regions for which the lubrication equations are expected to work.
To counter this difficulty, the calculation of $A_2$ (given in SI\cite{SI}) is performed in two steps:
First, the exact solution of the hydrodynamic problem in bipolar coordinates is examined in the flat-interface limit.
This solution provides the exact expressions of the angular velocity of the particle and the shear stress at the interface as a function of $\dot\gamma$ and $h_0$.
Second, the resulting expressions are fed into the classical lubrication framework to find the leading-order deformation of the interface and the downward force on the particle, from which the leading term is calculated for sufficiently small $C_a$ and $h_0$ as
\begin{equation}
\label{lubrication}
F_y/\sigma=-45\pi(2h_0)^{-1/2}C_a^2.
\end{equation}

\begin{figure}
    \begin{center}
        \includegraphics[width=0.9\columnwidth]{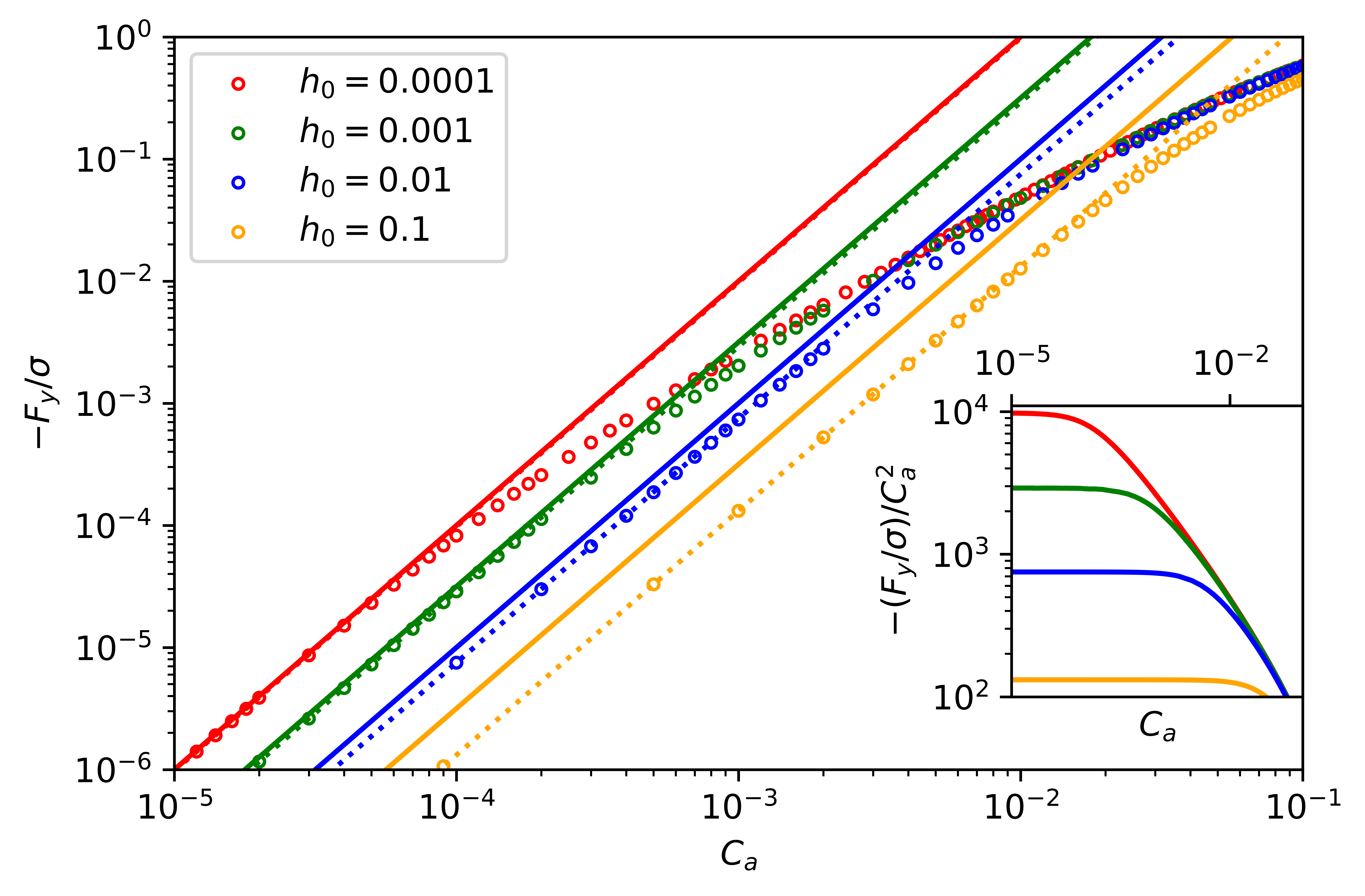}
        \caption{\label{fig:A2} Quadratic dependence of the indentation force on the shear rate. Symbols are numerical results, solid lines are eq. (\ref{lubrication}), dotted lines show the best $F_y\propto-C_a^2$ fit for $C_a\rightarrow 0^+$. Inset shows the saturation of $F_y/C_a^2$ in the $C_a\rightarrow 0$ limit (numerical results). $B_o=1$.
        }
    \end{center}
\end{figure}

\begin{figure}
    \begin{center}
        \includegraphics[width=0.9\columnwidth]{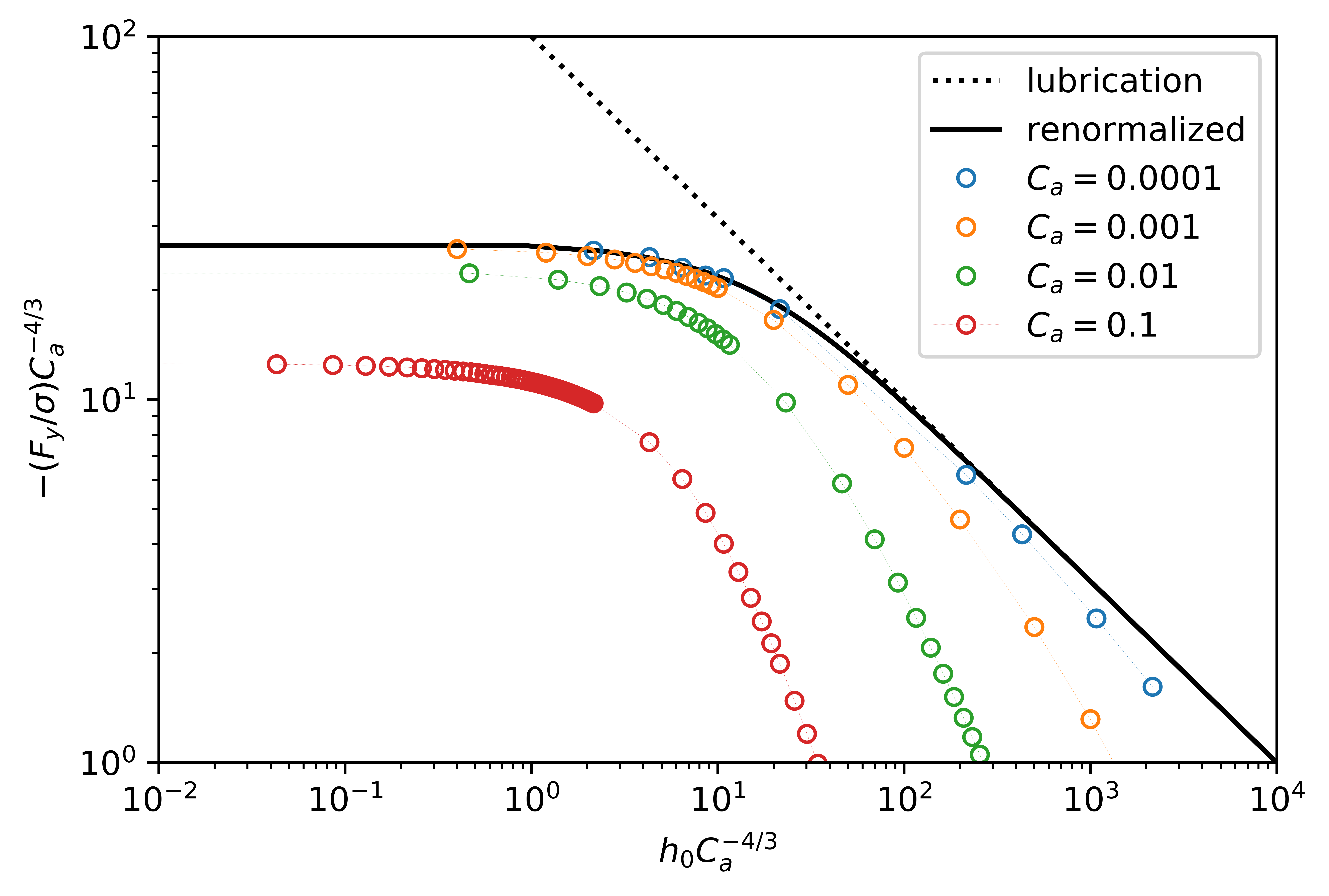}
        \caption{\label{fig:similarity}Numerical results (symbols) rescaled according to eq. (\ref{scaling}) with $\mu=\nu=4/3$. Dotted line refers to eq. (\ref{lubrication}). Solid line refers to eq. (\ref{renormalization}). $B_o=1$.
        }
    \end{center}
\end{figure}

The validity of the scaling (\ref{lubrication}) is tested in Fig. \ref{fig:similarity} showing that the lubrication approximation provides a correct approximation of $F_y$ for small enough $C_a$ but its range of validity decreases with decreasing $h_0$, shrinking to a point as $h_0$ tends to 0.
This puts an unexpectedly strong limitation on the range of applicability of the classical lubrication approach by imposing two contradictory constraints:
On the one hand, $h_0$ must be small enough for the higher-order terms in the $A_2$ expansion to be negligible, while on the other hand, the closer $h_0$ is to zero, the smaller is the range of $C_a$ for which expansion (\ref{Taylor}) can be used.
This problem arises because the radius of convergence of the series (\ref{Taylor}) decreases to 0 for $h_0\rightarrow 0^+$ due to the behavior of singularities in the complex plane and thus can not be solved by including higher-order terms in (\ref{Taylor}).
Indeed, the purely elastic indentation force is strictly zero for all $h_0>0$ and shows an analytic behavior on any interval of the $h_0<0$ subset.
The point $h_0=0$ thus represents a singular point at which the analytic nature of the elastic indentation curve breaks down.
Setting a small but finite value of $C_a$ regularizes the indentation curve making it analytic on sufficiently small intervals around any real $h_0$.
The singularity, however, does not disappear completely but is moved from the real axis in the complex plane, by a distance that scales as some positive power of $C_a$ (to be determined below), as discussed in detail, for example, in\cite{farutin2024}.
Equivalently, small but finite value of $h_0$ corresponds to a set of singularities in the complex $C_a$ domain, which all tend to 0 for $h_0\rightarrow 0^+$.
Since the singularity with the smallest absolute value sets the radius of convergence of the series (\ref{Taylor}), this radius tends to 0 as $h_0\rightarrow 0^+$, whence the divergence of $A_2$ for $h_0\rightarrow 0^+$.

The argument above suggests a theoretical approach that is valid beyond the lubrication approximation limit.
Since $C_a$ acts as a regularization parameter, rescaling $h_0$ by an appropriate power $\nu$ of $C_a$ would make the singularities in the complex plane of $h_0/C_a^\nu$ independent of $C_a$.
The resulting function $F_y$ would then scale as another power $\mu$ of $C_a$, independent of $h_0/C_a^\nu$.
This suggests the following scaling law for a given $B_o<\infty$:
\begin{equation}
\label{scaling}
F_y(h_0,C_a)/\sigma=C_a^\mu \mathcal F(h_0/C_a^\nu),
\end{equation}
where $\mathcal F$ is some function analytic around all real inputs.
The exponents $\mu$ and $\nu$ are found by matching the asymptotic scalings of $F_y$ for $h_0\rightarrow 0^+$ and $h_0\rightarrow 0^-$: 
For $\sigma>0$, $h_0<0$, and sufficiently small $C_a$, $F_y$ can be approximated by the elastic indentation law, which gives (as shown in SI\cite{SI})
\begin{equation}
\label{elastic}
F_y/\sigma=2B_o^{-1/2}h_0+O(h_0^2).
\end{equation}
Substituting (\ref{scaling}) into (\ref{lubrication}) and (\ref{elastic}), gives $\mu=\nu=4/3$ and $\mathcal F(\xi)\propto \xi^{-1/2}$ for real $\xi\rightarrow+\infty$ and $\mathcal F(\xi)\propto \xi$ for real $\xi\rightarrow-\infty$.
Setting $h_0=0$ in eq. (\ref{scaling}) suggests that $F_y\propto C_a^{4/3}$ for $h_0=0$, which is confirmed by comparison with numerical results (Fig. \ref{fig:similarity}).

Further progress is made by renormalization technique.
The steady state of the indentation problem satisfies a force balance condition, by which the downward force on the disc is balanced by an upward reaction force of the interface:
\begin{equation}
\label{balance}
F_y+\boldsymbol e_y\boldsymbol\cdot\int_{-L/2}^{L/2} \boldsymbol f_i(x) ds(x)=0,
\end{equation}
where $\boldsymbol f_i(x)$ is the sum of tractions applied by the interface on the fluids above and below.
This condition follows from the zero-flux condition in the $y$ direction across the interface.
Equation (\ref{balance}) is fundamental in showing that the interface deforms in order to balance the force applied on the particle regardless of $h_0$.
This means that even for $h_0>0$, the interface is indented in a way that is determined by its elasticity.
Most importantly, the indentation occurs due to the pressure in the lubrication region, and this region remains small assuming both $|h_0|$ and $C_a$ are sufficiently small but regardless of the relative smallness of $|h_0|$ and $C_a$.
The deformation of the interface, however, occurs on the capillary length scale $\lambda$.
The indented interface, therefore, can be approximated for small enough $|F_y|$ as flat on the length scale given by $\lambda$ and with its indentation depth obtained in the elastic indentation problem.
Inverting eq. (\ref{elastic}) yields $\delta h_0=-|F_y|B_o^{1/2}/2$.
The height of the particle above the indented interface therefore is equal to $h_0-\delta h_0$, which should be substituted in the denominator of eq. (\ref{lubrication}), yielding the following consistency relation:
\begin{equation}
\label{renormalization}
F_y/\sigma=-\frac{45\pi C^2_a}{\left(2h_0+|F_y|B_o^{1/2}/\sigma\right)^{1/2}}
\end{equation}
It is easy to check that eq. (\ref{renormalization}) provides the correct values of the exponents $\mu$ and $\nu$.
It also provides a much better approximation of the numerical results than the classical lubrication approximation, as shown in Fig.\ref{fig:similarity}.
Most remarkably, the renormalization model gives a finite value of the indentation force for $h_0=0$ and also gives reasonable results for $h_0<0$.

\begin{figure}
    \begin{center}
        \includegraphics[width=0.9\columnwidth]{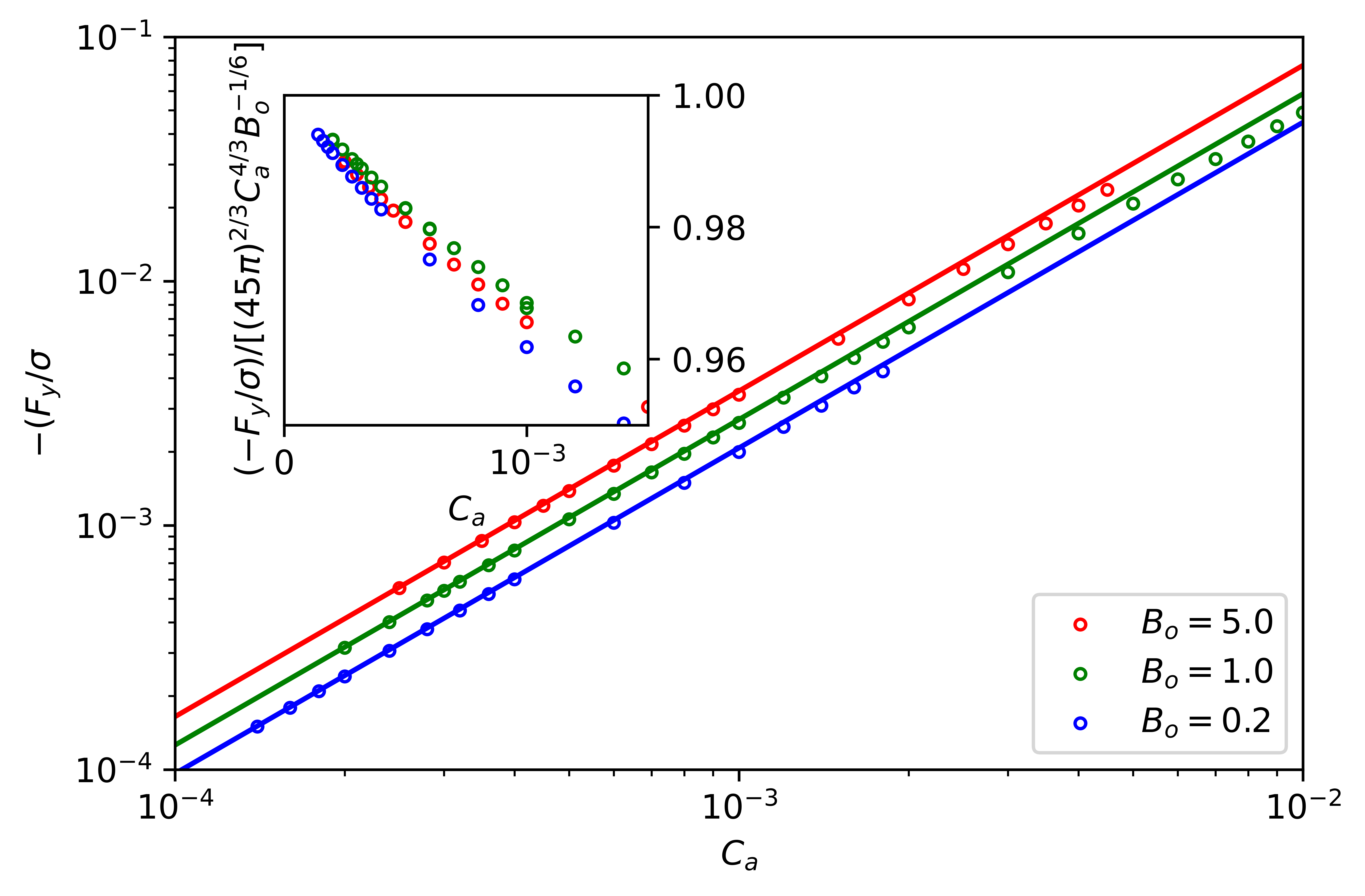}
        \caption{\label{fig:zerogap}Downward force as a function of $C_a$ for $h_0=0$ (symbols) compared with eq. (\ref{renormalization}) (lines) for several values of $B_o$. Inset shows that the ratio of two sides of eq. (\ref{renormalization}) obtained from numerical results as a function of $C_a$ for $h_0=0$ (same legend). The ratio tends to 1 with good precision. 
        }
    \end{center}
\end{figure}

Substituting $h_0=0$ into eq. (\ref{renormalization}), yields $F_y/\sigma=(45\pi)^{2/3}C_a^{4/3}B_o^{-1/6}$ in quantitative agreement with the numerical results, as shown in Fig. \ref{fig:zerogap}.
The highly non-linear behavior of the viscous indentation force beyond the classical lubrication limit can thus be extracted from the lubrication calculation and the purely elastic indentation problem.

\begin{figure*}
\begin{center}
\includegraphics[width=0.6\columnwidth]{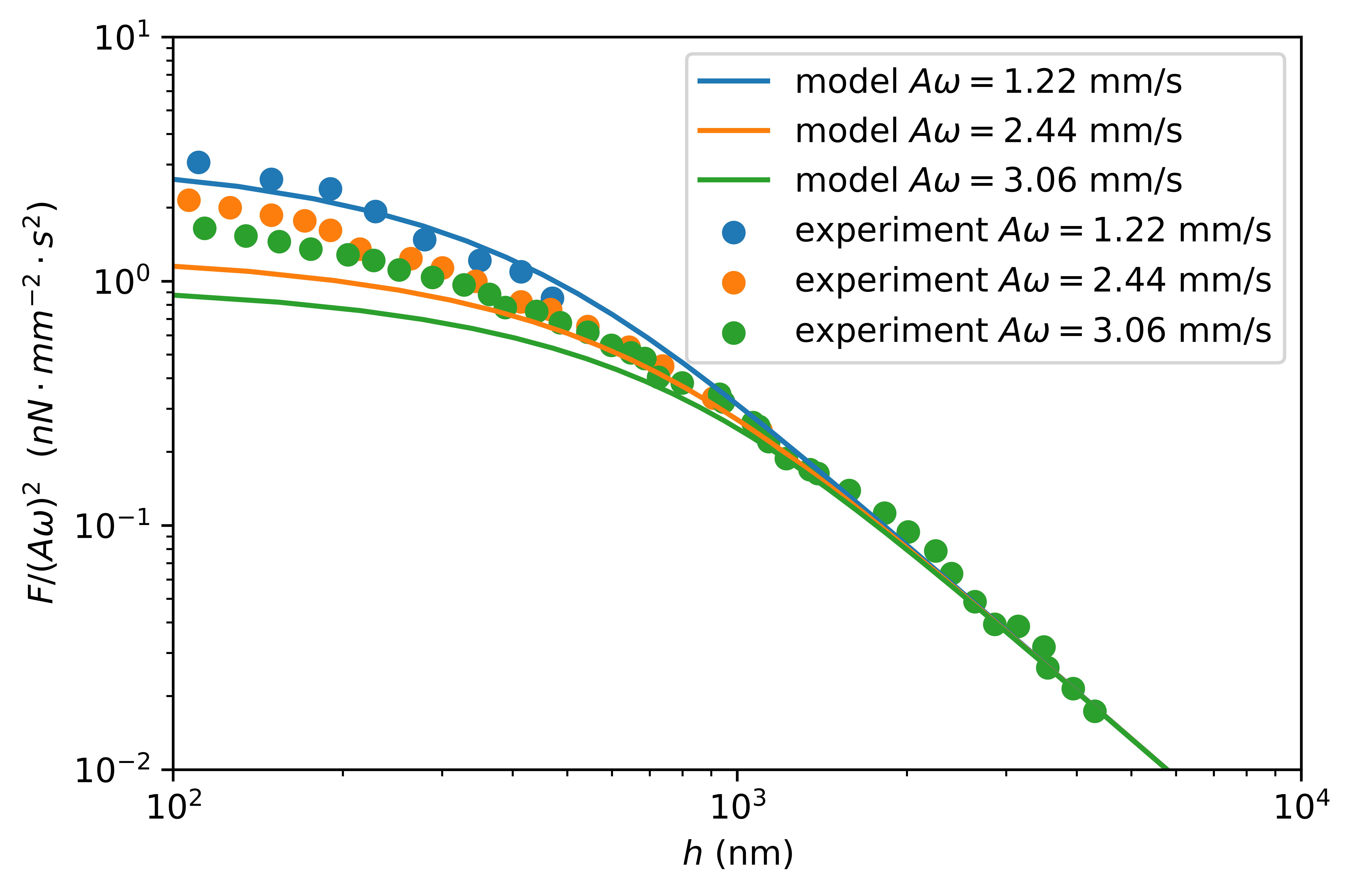}
\includegraphics[width=0.6\columnwidth]{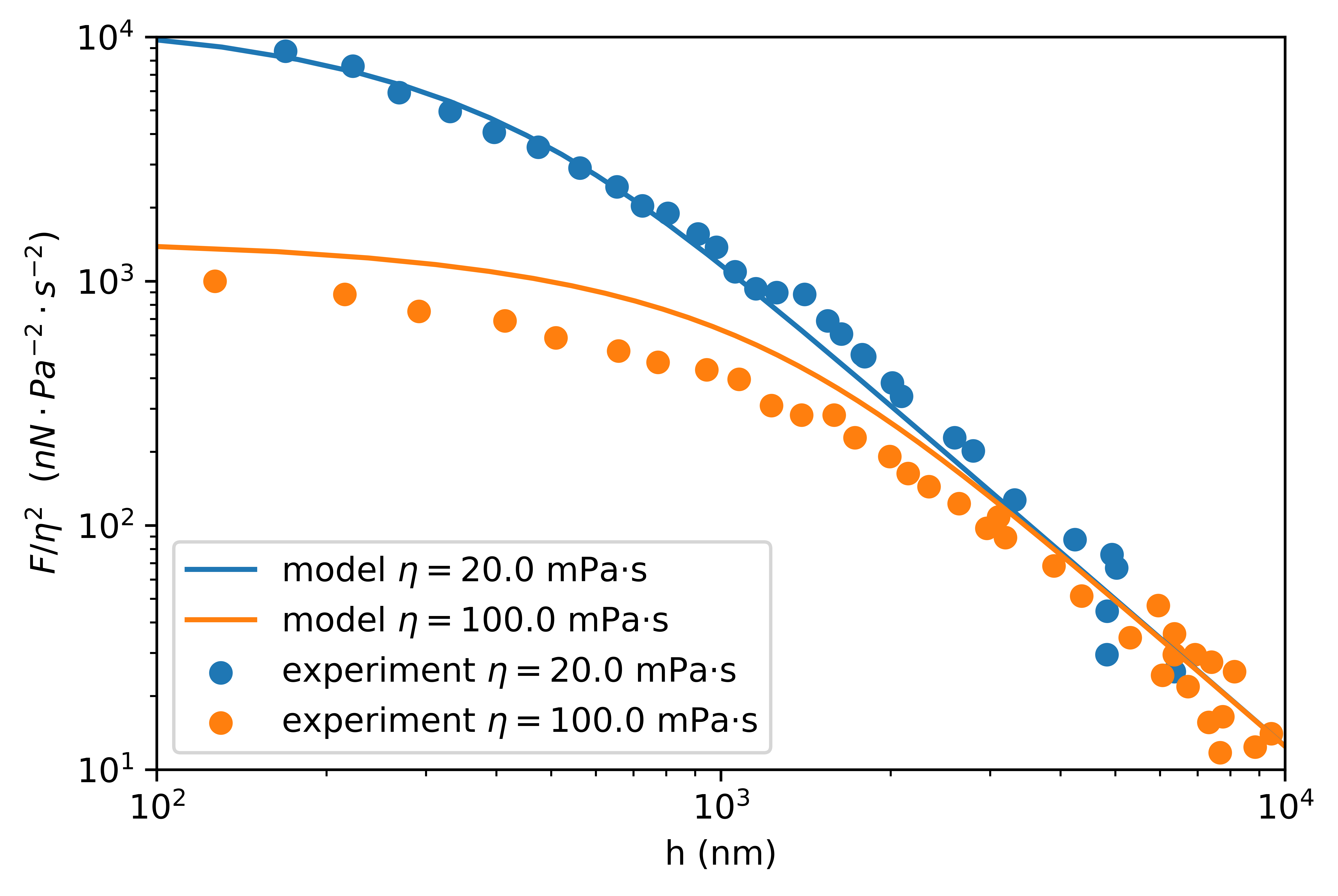}
\includegraphics[width=0.6\columnwidth]{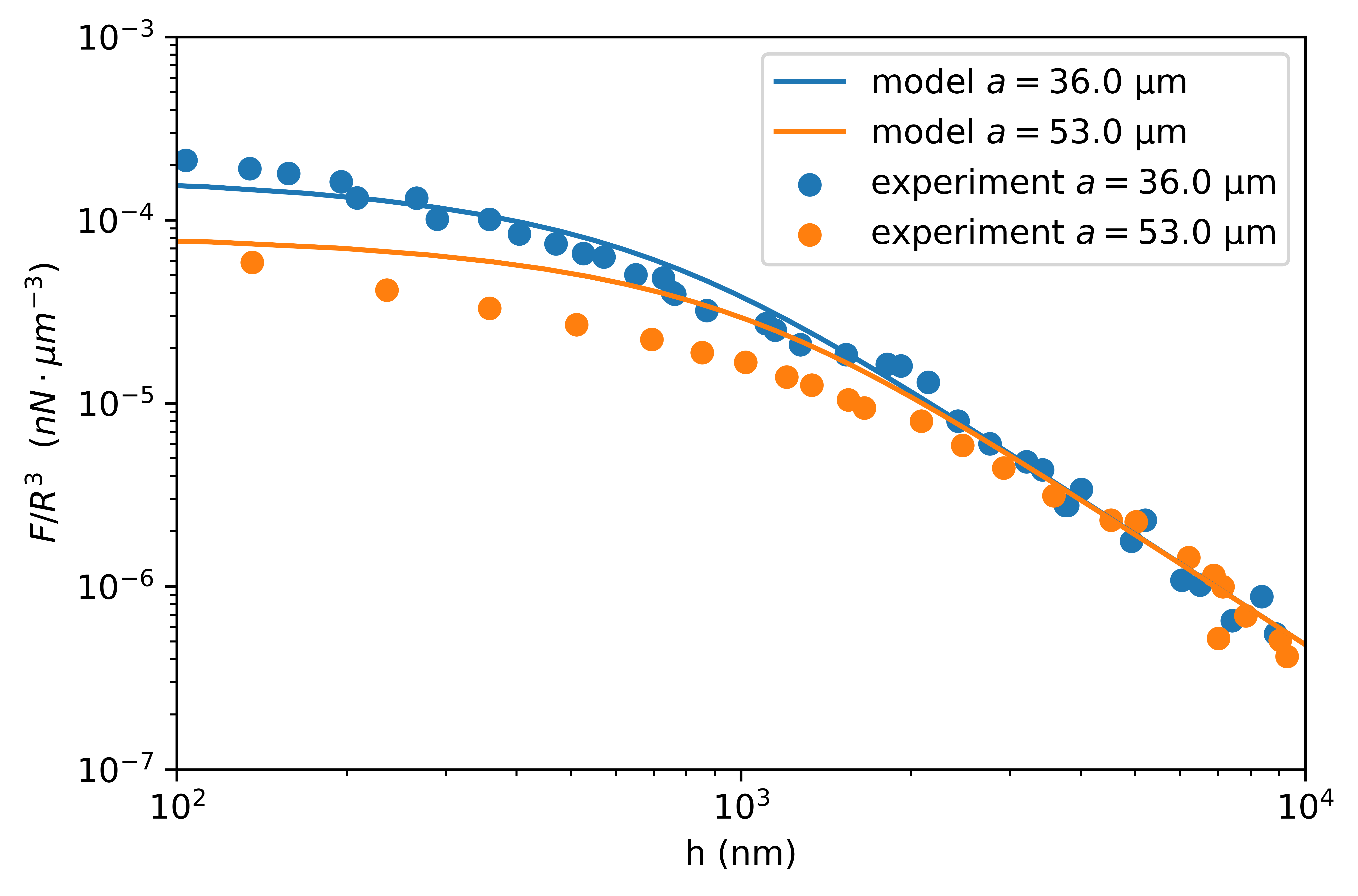}
\caption{\label{fig:experiment}Comparison of experimental results directly extracted from Fig. 3 of \cite{Zhang2025} (symbols) with the renormalization model (lines). All physical parameters are based on the experimental work. $\sigma=21mN/m$, $\Delta\rho g=2000 N/m^3$ (estimated), $a=36\mu m$ (unless stated otherwise), $\eta=20mPa\cdot s$ (unless stated otherwise). $A$ and $\omega$ are the amplitude and frequency of the stage oscillations, such that $\langle V^2\rangle=(A\omega)^2/2$ is the mean square velocity. $A\omega=1.22mm/s$ in central panel and $A\omega=2.59mm/s$ in right panel.}
\end{center}
\end{figure*}

The numerical results reported here concern a simplified 2D geometry but the renormalization approach is a general tool that can be used for a fully 3D problem.
It is therefore tempting to check whether the recent experimental measurements\cite{Zhang2025} of the indentation force can be interpreted within the renormalization framework.
The downward force in the classical lubrication approximation was calculated for a spherical particle moving with velocity $\boldsymbol V$ parallel to a fluid-fluid interface as\cite{Zhang2025}
\begin{equation}
\label{lift3D}
    F_y=-\frac{6\pi V^2}{25}\frac{\eta^2a^3}{\sigma d^2}.
\end{equation}
SI\cite{SI} contains the derivation of the elastic indentation force for a spherical particle, which has a complicated quasi-linear dependence on the indentation depth.
Combining the two equations together produces a renormalized relation between $d$ and $F_y$, which is solved numerically and compared with the experimental data\cite{Zhang2025} in Fig.\ref{fig:experiment}.

As already highlighted in\cite{Zhang2025}, the experimental results agree well with the classical lubrication law (\ref{lift3D}) for large enough $d$ but deviate from the $F_y\sim d^{-2}$ power law for smaller $d$.
Figure \ref{fig:experiment} shows that this deviation can be captured quite well by the renormalization model.
In particular, the agreement is quantitative for the smallest velocity amplitudes used in the experiment.
For larger velocity amplitudes the discrepancy becomes quite noticeable which suggests that the leading-order approximation becomes only qualitatively correct in this case.

It must be noted that the renormalization model does not use any phenomenological parameters.
The only value which was not taken directly from the experiment was the density difference between the upper (silicon oil) and the lower (glycerol) fluids, which is estimated to be equal to 200$kg/m^3$ in the renormalization model.
However, due to the large capillary length in the experiment, the density difference enters the $F_y(d)$ expression only as a logarithmic correction and changing its value in a wide range does not produce a visible change of the renormalization curves in Fig. \ref{fig:experiment}.

\paragraph{Conclusions}
This work presents a detailed study of the contactless indentation of a fluid-fluid interface by a rigid particle under flow.
It is found that the indentation force remains finite as the height of the particle above the undeformed interface becomes zero or even goes negative, contrary to predictions of the classical lubrication theory.
It is further found that the particle remains separated from the interface by a fluid film of a well-defined thickness even when the height of the particle above the equilibrium level of the interface is negative.
The gap renormalization model, proposed here, combines the classical lubrication calculation with the purely elastic indentation law of the substrate.
This model provides the exact values of both the exponent and the constant of the leading-order scaling of the indentation force with the shear rate when the height of the particle above the undeformed interface tends to zero.
The model also explains the reasons of the breakdown of the classical lubrication approximation when the particle is too close to the interface.
Applying the height-renormalization technique to the 3D problem provides a quantitative interpretation of the experimental results.

Most of the results presented above pertain to $B_o=1$.
The main conclusions of this study remain valid for other values of $B_o$, except for the limit $B_o\rightarrow\infty$, which corresponds to $\sigma=0$.
The problem remains well-posed for $\sigma=0$, with gravity acting as the sole elastic force.
Despite many qualitative similarities, the scaling laws are different for $\sigma=0$.
For example, the elastic force scales as $|h_0|^{3/2}$ for $h_0<0$ and the downward force in the classical lubrication approximation scales as $\dot\gamma^2h_0^{-3/2}$ for $h_0>0$ (both scalings are derived in SI\cite{SI}).
This leads to completely different exponents $\mu=1$ and $\nu=2/3$ in eq. (\ref{scaling}).
Furthermore, the gap renormalization technique gives the correct value only for the exponent but not for the constant of the downward force as a function of $\dot\gamma$ for $h_0=0$ (although the constant is off by less than 20\%).
This is expected since the interface is not flat in the lubrication region due to the local nature of the gravity force.

The particle is subject to a downward force which keeps it at a certain height in this work.
A related problem is calculating the migration velocity of a force-free particle.
These two problems are similar for $h_0>0$ and small enough $C_a$ but should be completely different otherwise.
Indeed, it is shown here that the downward force acting on the particle is essential in indenting the inteface.
Consequently, there should be no well-defined migration velocity for $h_0<0$ in the force-free migration:
Starting from $h<0$, the visco-elastic relaxation of the indented interface would propel the particle upwards on the time scale of $\tau_c$ regardless of $C_a$.
Furthermore, it is not possible to compute the quasi-static migration velocity by neglecting the migration and letting the interface relax to a stationary shape for given $h_0$:
Since such solutions are only an approximation of the migration, the particle would eventually collide with the interface in this procedure, if either $h_0<0$ or $h_0\ge 0$ and $C_a$ is sufficiently large.

This study shows that the classical lubrication models, that are traditionally employed to find the lift force or migration velocity for a particle near a deformable substrate, are valid only in the weak-flow limit and give qualitatively incorrect prediction outside of their validity limit.
The method for deriving the scaling laws outside of the weak-flow limit and the renormalization approach presented here can be applied to other types of deformable substrates, giving qualitatively or even quantitatively correct results outside of the range of validity of the classical lubrication calculations.
The numerical method developed for this study remains stable and precise even when the fluid film between the particle and the interface is 1000 times thinner than the particle radius, while simultaneously resolving the flows on the length-scales much larger than the particle.
With straightforward modifications, this method can be applied to a diverse set of visco-elastic substrates.
Finally, the exact solution of the problem for a flat interface, given in this work, can be used for other problems in which the fluid stress varies on the length scale comparable to the particle size.
Using this solution can extend the set of problems in which the lubrication approximation gives an exact expression for the downward force or migration velocity.

\paragraph{Acknowledgments}

The author thanks CNES (Centre National d’Etudes Spatiales) and the French-German university program "Living Fluids" (grant CDFA-Q1-14).
The author thanks Dalei Jing, Yi Sui, Abdelhamid Maali and Chaouqi Misbah for many stimulating discussions of particle migration near a deformable boundary.
The calculations were performed on the cactus2 cluster of the GRICAD infrastructure and the author thanks Vikhram Duffour and Philippe Beys who maintain the cluster.

\bibliographystyle{apsrev}

\bibliography{ref}

\onecolumngrid
\section{Supplementary Information}
\subsection{Full model and numerical method}
\subsubsection{Full model}

Since inertia effects are neglected, the flow in fluids above and below the interface is described by the Stokes equations
\begin{equation}
	\label{Stokes}
	\eta\nabla^2\boldsymbol u-\boldsymbol\nabla p=0,\,\,\,\boldsymbol\nabla\boldsymbol\cdot\boldsymbol u=0,
\end{equation}
where $\boldsymbol u$ and $p$ are the fluid velocity and the pressure, respectively.
Equation (\ref{Stokes}) is supplemented by the following boundary conditions:
\begin{enumerate}
	\item Periodic boundary condition in $x$ direction: $\boldsymbol u(x+L)=\boldsymbol u(x)$, $p(x+L)=p(x)$, where $L$ is the period.
	\item Shear flow at large $|y|$: $\boldsymbol u(x,y)-\dot\gamma\boldsymbol e^x y=0$ for $y\rightarrow\pm\infty$.
	\item No-slip boundary condition at the particle boundary: $\boldsymbol u(\boldsymbol r)=\boldsymbol V+\Omega\boldsymbol e^z\times(\boldsymbol r-\boldsymbol r_c)$ for $|\boldsymbol r-\boldsymbol r_c|=a$, where $V$ and $\Omega$ are the translational and rotational velocities of the particle, respectively, and $\boldsymbol r_c=(x_c,y_c,0)$ is the center of mass of the particle.
	\item No-slip boundary condition at the interface, which dictates that $\boldsymbol u(x,y)=\boldsymbol u_i(x,y_i(x))$ for $y\rightarrow y_i(x)^\pm$ for any $x$, where $\boldsymbol u_i(x)$ and $y_i(x)$ are the velocity and $y$ position of the interface at $x$.
	\item The force balance at the interface which dictates that $(\lim\limits_{y\rightarrow y_i(x)^+}\mathsf s-\lim\limits_{y\rightarrow y_i(x)^-}\mathsf s)\boldsymbol\cdot\boldsymbol n_i(x)+\boldsymbol f_i(x)=0$ for any $x$,
	where $\mathsf s=\eta\left[\boldsymbol\nabla\otimes\boldsymbol u+(\boldsymbol\nabla\otimes\boldsymbol u)\right]-p\mathsf I$, is the stress tensor of the fluid and $\boldsymbol n_i(x)$ and $\boldsymbol f_i(x)$ are the upward normal and the density of interfacial forces at given $x$, respectively.  Here $\mathsf I$ is the metric tensor of the Cartesian coordinates.
\end{enumerate}

The density of interfacial forces is written as 
\begin{equation}
	\label{force}
	\boldsymbol f_i(x)=\left[\sigma c_i(x)-\Delta\rho g y_i(x)+\Delta p_0\right]\boldsymbol n_i(x),
\end{equation}
where $c_i(x)$ is the curvature of the interface (positive when $y_i''>0$) and $\Delta p_0$ is the hydrostatic pressure jump across the interface.

The forces applied by the particle on the surrounding fluid $\boldsymbol f_d=-\mathsf s\boldsymbol\cdot\boldsymbol n_d$, where $\boldsymbol n_d$ is the outward normal to the particle, are unknown but must add up to the downward force, while satisfying the zero-torque condition:
\begin{equation}
\label{totalForce}
	\oint \boldsymbol f_d(l)dl=\boldsymbol F,\,\,\, \oint \boldsymbol f_d(l)\times [\boldsymbol r_d(l)-\boldsymbol r_c] dl=0,
\end{equation}
where $l$ is the arc-length coordinate on the particle boundary and $\boldsymbol r_d(l)$ is the position vector for given $l$.

\subsubsection{Numerical method}
The problem is solved using the boundary integral method\cite{Pozrikidis92}, which combines the Stokes equations and the boundary conditions 1 to 5 into one equation
\begin{equation}
	\label{BIE}
	\boldsymbol u(\boldsymbol r)=\boldsymbol u^\infty(\boldsymbol r)+\boldsymbol u^i(\boldsymbol r)+\boldsymbol u^d(\boldsymbol r)=
	\boldsymbol u^\infty(\boldsymbol r)+\frac{1}{\eta}\int_{-L/2}^{L/2} \mathsf G^p(\boldsymbol r-\boldsymbol r_i(x))\boldsymbol\cdot\boldsymbol f_i(x)\frac{dl_i}{dx}dx
	+\frac{1}{\eta}\oint \mathsf G^p(\boldsymbol r-\boldsymbol r_d(l))\boldsymbol\cdot\boldsymbol f_d(l)dl,
\end{equation}
where $\boldsymbol u^i$ and $\boldsymbol u^d$ are the contributions of the interface and the particle  to the flow, $l_i$ is the arc length on the interface, and $\mathsf G^p$ is the Green's function for the Stokes equation (\ref{Stokes}) and periodic boundary conditions in two dimensions, the explicit form of which is listed in\cite{Pozrikidis92}.

In order to ensure high precision of the numerical solution, spectral parametrization is used for all scalar and vector fields on the interface and the particle boundary.
The interface is parametrized by its height $y_i(\xi_i)$, where the reference variable $\xi_i\in[0,1]$ is chosen by the relation
\begin{equation}
\label{interface_parametrization}
\frac{dx}{d\xi_i}=L\left[1+\alpha_i\tanh \frac{\cos(2\pi\xi_i)}{\epsilon_i}\right],
\end{equation}
where $\alpha_i\in[0,1)$ is a constant which sets the refinement of the mesh near the particle and $\epsilon_i$ is the regularization parameter that sets the width of the transitional region between the fine and the coarse regions of the mesh.
The parametrization (\ref{interface_parametrization}) maps a homogeneous mesh in the $\xi_i$ domain onto the interface in such a way, that about half of the mesh points [region $(1/4,3/4)$] are mapped to a small interval of $-L(1-\alpha_i)/4<x<L(1-\alpha_i)/4$ near the particle.
The length of this interval is chosen from about $3a$ for $h\simeq -1$ to about $0.25a$ for $h\simeq 0$ by adjusting $\alpha_i$.
Quasi-homogeneous parametrization is preferable for $h>1$.
The interfacial geometry and force are computed in a straightforward way by expanding $y_i$ into a Fourier series of $\xi_i$ thanks to the periodicity relation $y_i(\xi_i+1)=y_i(\xi_i)$.
Given a shape of the interface, the pressure difference across the interface is computed to satisfy the zero sum of total force applied by the interface and the particle on the fluid, which is necessary for the vertical flux to be equal to zero, as discussed in the main text.
The hydrodynamic self-interaction of the interface is computed according to the eq. (\ref{BIE}) using the singularity subtraction and refined meshes to improve the precision of the calculation\cite{Farutin2014b}.
Typically, 256 harmonics are used for time-resolved simulations and up to 2048 harmonics are used for steady-state calculations  when high precision is required.

Homogeneous parametrization is used for the disk:
\begin{equation}
	\label{fdisk}
	f_{d,x}(l)+if_{d,y}(l)=\sum_{k=-k_{max}}^{k_{max}}f_{d,k}e^{ikl/a},\,\,\,u_x(l)+iu_y(l)=\sum_{k=-k_{max}}^{k_{max}}u_{d,k}e^{ikl/a},
\end{equation}
where $f_{d,k}$ and $u_k$ are complex amplitudes.
The Fourier series expansion for the velocity field is applied to $\boldsymbol u^\infty$, $\boldsymbol u^i$ and $\boldsymbol u^d$.

Precise calculation of the hydrodynamic interactions in eq. (\ref{BIE}) is a challenging task when the distance between the particle and the interface is much smaller than the particle size.
This challenge is addressed in the present study by decomposing the Green's function into two parts: $\mathsf G^p(\boldsymbol r)=\mathsf G^\infty(\boldsymbol r)+\delta \mathsf G^p(\boldsymbol r)$, where $\mathsf G^\infty(\boldsymbol r)$ is the Green's function of an infinite domain and $\delta \mathsf G^p(\boldsymbol r)$ is the correction due to the periodic boundary conditions.
The correction due to $\delta\mathsf G^p$ for the velocity fields at the disk is smooth and is computed in real space using a regular mesh, then transformed in Fourier series of $l$.
Typically, 64 points is sufficient unless the disk diameter is very close to $L$.
The singular contribution $\mathsf G^\infty$ for a point force $\boldsymbol f$ located at a point $\boldsymbol r$ is computed directly in the Fourier representation, using explicit formulae:
\begin{equation}
    \label{GreenFourier}
    \frac{1}{2\pi a}\int e^{-ikl/a}(\boldsymbol e^x+i\boldsymbol e^y)\boldsymbol\cdot \mathsf G^\infty\left[\boldsymbol r-\boldsymbol r(l)\right]\boldsymbol\cdot\boldsymbol fdl= \left[(f_x+if_y)G^1_k(\boldsymbol r-\boldsymbol r_c)+(f_x-if_y)G^2_k(\boldsymbol r-\boldsymbol r_c)\right],
\end{equation}
where $G^1_k$ and $G^2_k$ are functions of $\boldsymbol r-\boldsymbol r_c$ that are computed explicitly for all $k$.
The Fourier harmonics $u_k$ can then be computed as integrals of $(f_x+if_y)G^1_k$ and $(f_x-if_y)G^2_k$ over the corresponding boundary.
The same approach for two parallel walls is discussed in detail in \cite{Abbasi2022}.
The main advantage of this method to compute the velocity field at the boundary of the particle is that the kernels $G^1_k$ and $G^2_k$ are regular for $|\boldsymbol r-\boldsymbol r_c|=a$ and thus can be integrated along the interface by conventional methods.
For large $k$, computing the integrals of $G^1_k$ and $G^2_k$ along the interface requires a mesh on the interface that is sufficiently fine to resolve the wave-lengths of order $2\pi a/k$ but in practice, the amplitudes of high-order harmonics decay exponentially with $k$, with attenuation that usually scales as a square root of the minimum distance between the particle and the interface.

Using the computed Fourier harmonics of the velocity field $\boldsymbol u^i$ and the imposed flow $\boldsymbol u^\infty$, the forces at the disc boundary, the angular velocity and the translation velocity of the disk are found by solving the no-slip boundary condition at the disk, together with the zero-torque condition and the given downward force directly in the Fourier representation.
For a disk in an infinite fluid, the Green's kernel $\mathsf G^\infty$ is diagonal in the Fourier representation but the periodic boundary conditions add a small correction which also contains off-diagonal terms.
The resulting matrix is then inverted numerically and stored since it does not change over the course of the simulation. 
Once the force at the disk boundary is computed in the Fourier representation, the resulting velocity $\boldsymbol u^d$ is computed directly for the $\mathsf G^\infty$ part using expressions similar to (\ref{GreenFourier}), while the correction due to the periodic boundary conditions $\delta\mathsf G^p$ is integrated in Cartesian coordinates using a small number of homogeneously distributed points on the disk boundary.

\begin{figure}
    \begin{center}
        \includegraphics[width=0.3\textwidth]{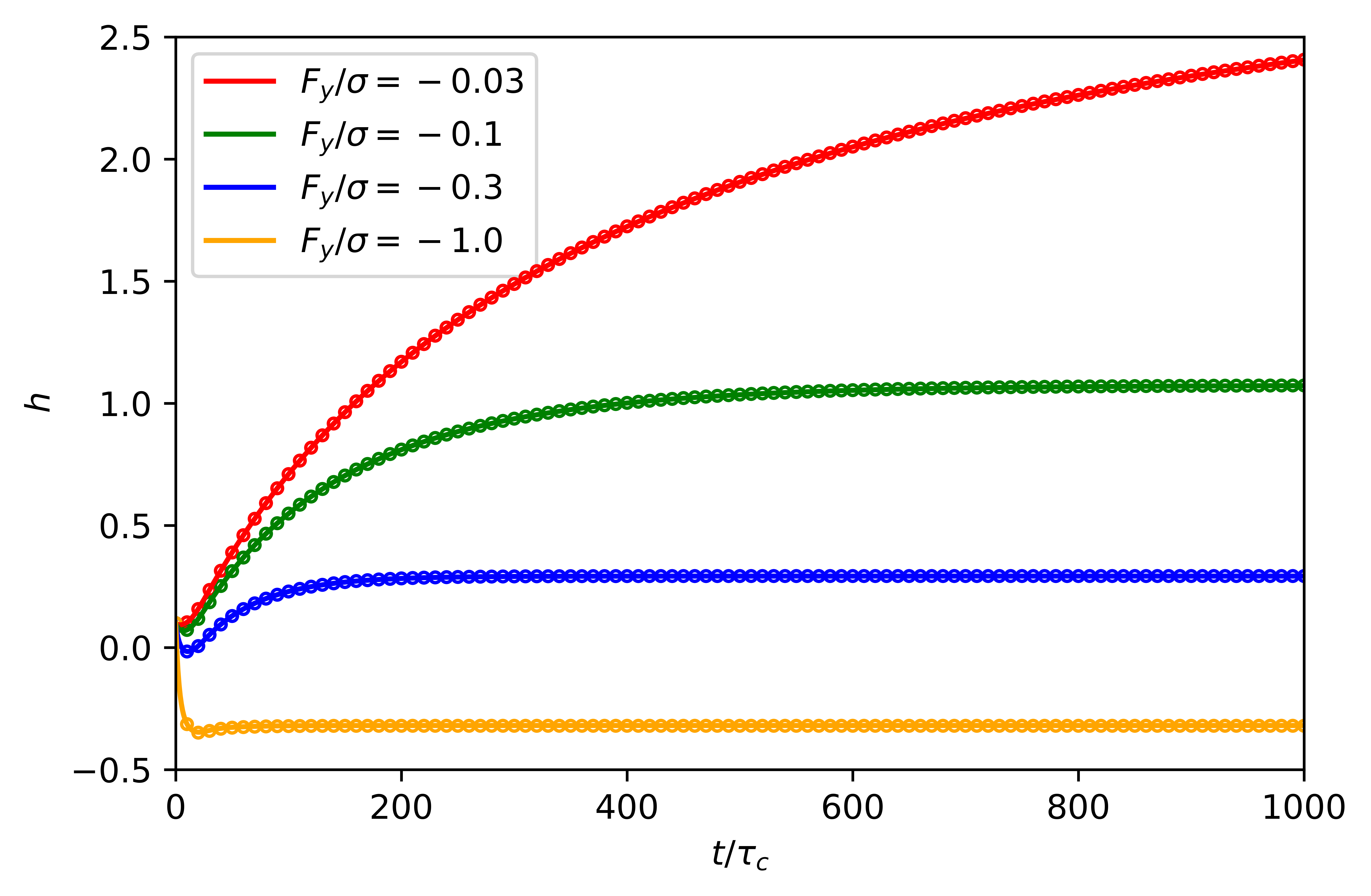}
        \includegraphics[width=0.3\textwidth]{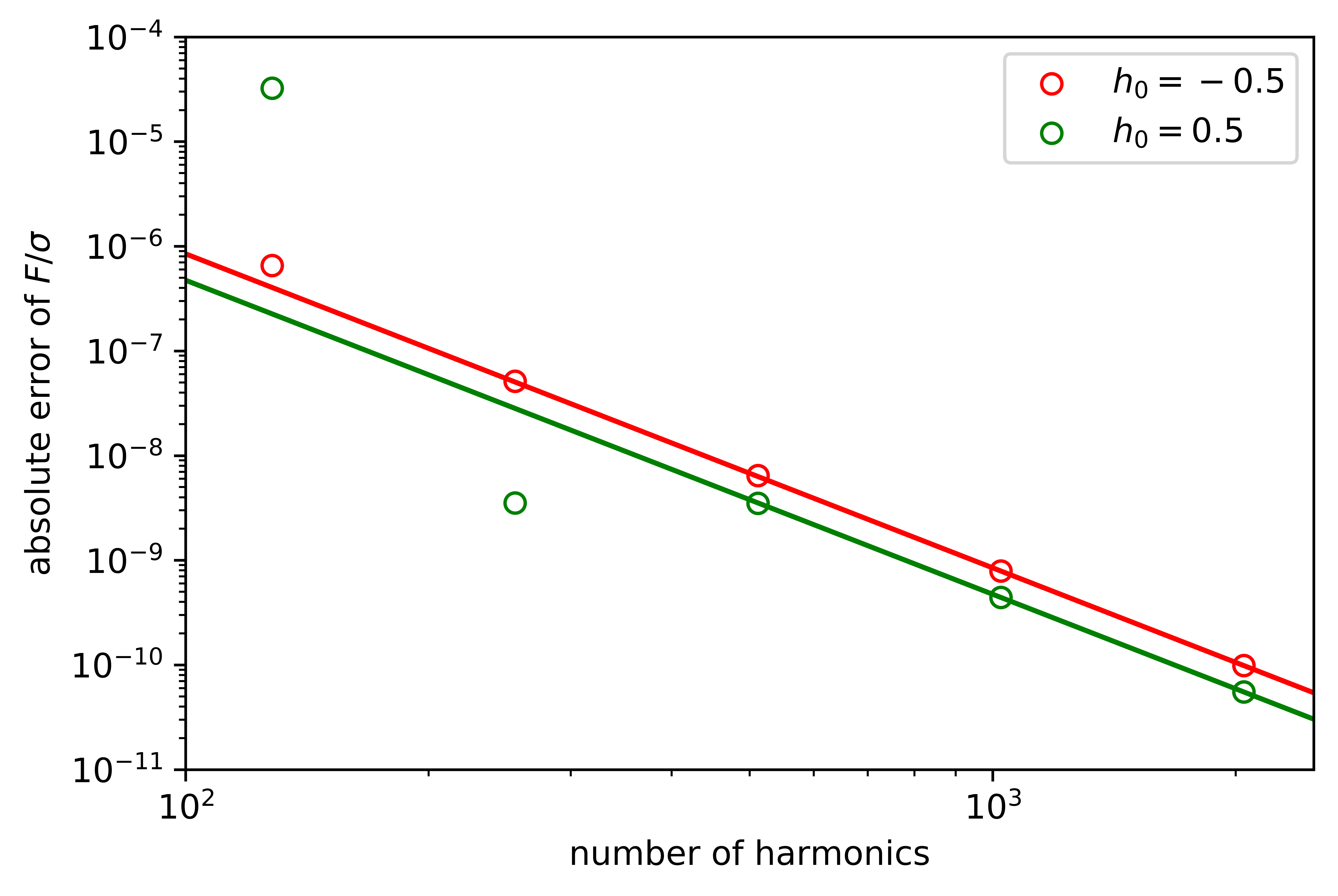}
        \includegraphics[width=0.3\textwidth]{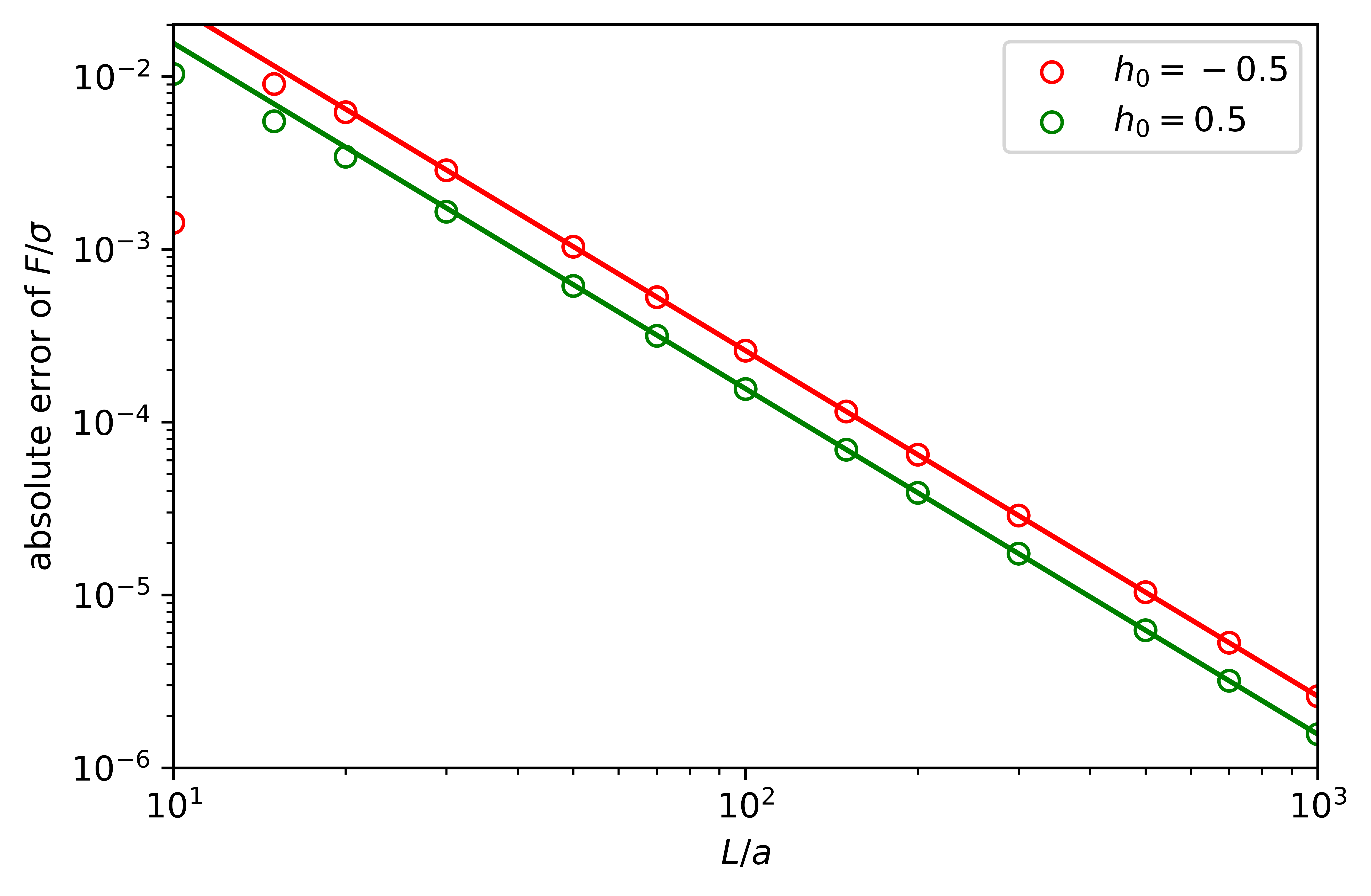}
        \caption{\label{fig:validation}Left: Comparison of the present method (curves) with a general-purpose method (symbols) for time-dependent indentation of the interface. $C_a=0.1$, $B_o=0.5$, $L=20a$, 256 harmonics used for interface parametrization. The maximum absolute difference for $h(t)$ between the two methods is about $0.001$ and is mostly due to the simple penalization method of imposing the no-slip boundary condition at the particle boundary that is used in the general-purpose method. Time step is $0.01\tau_c$. Center: Convergence of $F_y$ for given $h_0$ with the number of harmonics $N_h$ used for interface parametrization at steady state. Symbols are numerical results, lines are best fit with error$\propto N_h^{-3}$. $C_a=0.1$, $L=1000a$. Right: Convergence of $F_y$ for given $h_0$ with $L/a\rightarrow\infty$ at steady state. Symbols are numerical results, lines are best fit with error$\propto L^{-2}$. $C_a=0.1$, $B_o=1$, 2048 harmonics  used for interface parametrization.}
    \end{center}
\end{figure}

After adding together all contributions to the interface velocity, the interface shape and the position of the particle are updated in the co-coming frame as
\begin{equation}
    \label{update}
    \frac{dx_c}{dt}=0\,\,\,\frac{dy_c}{dt}=v_{0,y},\,\,\,\frac{dy_i(x)}{dt}=\frac{[\boldsymbol u_i(x)-u_{0,x}\boldsymbol e^x]\boldsymbol\cdot\boldsymbol n_i(x)}{\boldsymbol n_i(x)\boldsymbol\cdot\boldsymbol e^y}.
\end{equation}
There are two options for time-dependent simulations, either the downward force is fixed in time and the particle is free to migrate, or the height of the particle is fixed and the downward force is adjusted at each time step to keep the vertical velocity of the particle equal to zero, as the interface evolves to a steady-state solution.
The second option is preferable for measuring the lift force as a function of $y_c$.

In practice, the lift force is found by looking for a steady-state solution of the problem at a given particle height using a non-linear solver, which solves the regularized linearized problem over several iterations.
For an indented interface and weak flows, it is often beneficial to generate an initial approximation of the steady-state solution using the time-dependent solution, to avoid crossing between the particle and the interface.
A hydrodynamic solver is used to find the normal force at the interface that leads to a purely tangential flow in the co-moving frame for given interface shape and particle position.
This solver is used to validate the exact solution for a flat interface presented below.
The flow field in the fluids was calculated from eq. (\ref{BIE}) using singularity subtraction and refined meshes for $\boldsymbol u^i$ and the calculation in Fourier representation for $\boldsymbol u^d$ in order to avoid loss of precision for points near the interface or the particle boundary.

Each part of the numerical procedure was validated separately by testing the numerical results for problems that are simple enough to have an analytical solution (such as an image-based Green's function for Stokes equation in presence of a rigid disk).
The full method was validated by comparing the results with an independent general-purpose boundary integral method code for a time-dependent solution of the problem and the terminal position of the particle (Fig.\ref{fig:validation}, left).
The numerical method shows 3rd-order convergence with the number of harmonics used for interface discretization (Fig.\ref{fig:validation}, center), which is due to the boundary integral calculation, which remains weakly singular even with singularity subtraction.
The convergence with $L$ is of 2nd order (Fig.\ref{fig:validation}, right).
Note that fixing the average height of the interface instead of its position at $x=\pm L/2$ decreases the convergence with $L$ to 1st order.

$L$ was set to $1000a$ in all cases.
Triple mesh refinement was used in most cases (8 times more points on the finest mesh).
Starting with 128 to 512 harmonics the initial approximation for the steady-state solver was found.
The number of shape harmonics was then doubled repeatedly until reaching 2048 to find better approximations of the steady state.
The number of the disk force harmonics was similarly doubled at each step.
With this setup, from several hours to a day are needed to find the steady-state solution of the problem for one set of parameters for fluid film thickness above $0.001a$, depending on the convergence of the non-linear solver.
The non-linear solver usually fails to converge for lower film thicknesses due to finite precision of the floating-point calculation.
Changing double precision to extended (long double) was usually sufficient to study fluid film thickness as small as $0.0001a$ but the calculations took several days for one set of parameters.
In all cases, the precision of the lift force calculation was estimated by comparing the results for 1024 and 2048 shape harmonics, for which the variation was usually insignificant.
Nevertheless, a noticeable variation was observed when the fluid film thickness was approaching the applicability limit of the method.

\subsection{Elastic indentation in 2D}

\begin{figure}
    \begin{center}
        \includegraphics[width=0.8\textwidth]{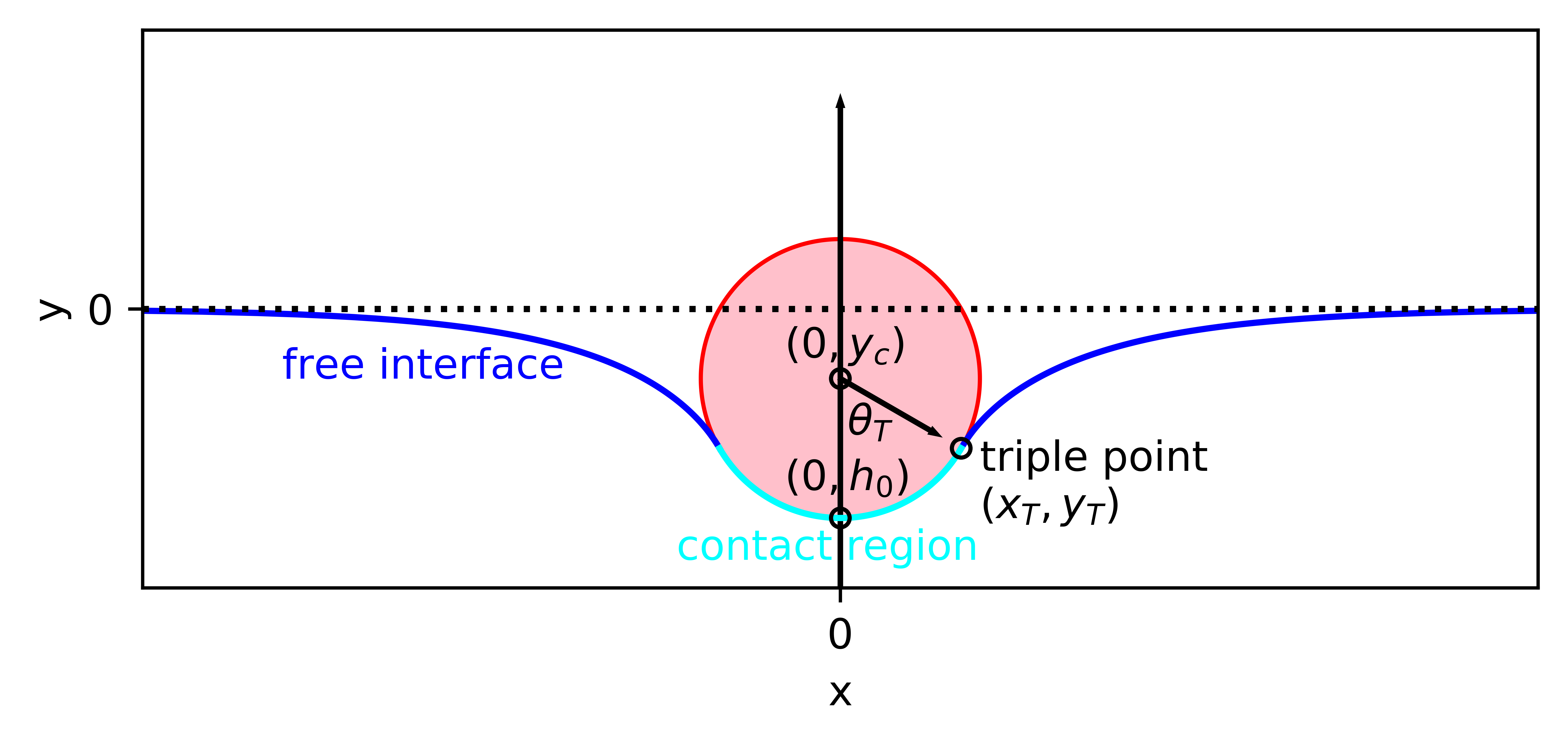}
        \caption{\label{fig:elastic}Schematic drawing of the elastic indentation problem.}
    \end{center}
\end{figure}
The problem setup is shown in Fig.\ref{fig:elastic}.
The interface for $x>0$ consists of two parts: the part where the interface touches the circle and the remaining free part.
The angle between the downward direction and the vector pointing to the triple point of the sphere and the two fluids is denoted as $\theta_T\in(0,\pi/2)$.
With this parametrization, the triple point $(x_T,y_T)$ is given by coordinates $(a\sin\theta_T,a(1+h_0-\cos\theta_T))$.

The free part of the interface obeys the following equation:
\begin{equation}
    \label{elastic2D}
    \frac{\sigma y_i''(x)}{(1+y_i'(x)^2)^{3/2}}-\Delta\rho g y_i(x)=0,
\end{equation}
where the first term describes the tension force and the second term describes the gravity force in the Monge parametrization.
This equation needs to be solved with boundary condition 
\begin{equation}
    \label{bc2D1}
    y_i(\infty)=0
\end{equation}
at infinity and the tangent direction at the triple point 
\begin{equation}
    \label{bc2D2}
    y_i'(a\sin\theta_T)=\tan\theta_T.
\end{equation}

Solving eq. (\ref{elastic2D}) with boundary condition (\ref{bc2D1}) yields an expression of $x$ as an explicit function of $y_i$
\begin{equation}
    \label{interface2D}
    x=\lambda \left[- \sqrt{4 - \frac{y_i^{2}}{\lambda^{2}}} + \operatorname{atanh}{\left(\sqrt{1 - \frac{y_i^{2}}{4 \lambda^{2}}} \right)}\right] + x_{0}
\end{equation}
where $x_0$ is an integration constant.

The solutions of the problem are parametrized by a variable $\xi_T$, such that
\begin{equation}
    \label{yT}
    y_T=-\frac{\sqrt{2} \lambda \xi_T}{\xi_T + 1}
\end{equation}
Since according to (\ref{interface2D}) $y_T$ can not be less that $-\sqrt{2}\lambda,$ the parametrization (\ref{yT}) is chosen to map the domain $\xi_T\in(0,\infty)$ to the the possible range $y_T\in(-\sqrt{2}\lambda,0)$.

Boundary condition (\ref{bc2D2}) then gives
\begin{equation}
    \label{theta2D}
    \theta_T=\operatorname{atan}{\left(\frac{\xi_T \sqrt{\xi_T^{2} + 4 \xi_T + 2}}{2 \xi_T + 1} \right)}
\end{equation}
The $x$ coordinate of the triple point is then calculated as
\begin{equation}
    \label{xT}
    x_T=a\sin\theta_T=\frac{a \xi_T \sqrt{\xi_T^{2} + 4 \xi_T + 2}}{\left(\xi_T + 1\right)^{2}}.
\end{equation}
Applying the relation (\ref{elastic2D}) to $x_T$ and $y_T$, yields an expression for $x_0$:
\begin{equation}
    \label{x02D}
    x_0=- \lambda \operatorname{atanh}{\left(\frac{\sqrt{2} \sqrt{\xi_T^{2} + 4 \xi_T + 2}}{2 \left(\xi_T + 1\right)} \right)} + \frac{\sqrt{\xi_T^{2} + 4 \xi_T + 2} \left(a \xi_T + \sqrt{2} \lambda \xi_T + \sqrt{2} \lambda\right)}{\left(\xi_T + 1\right)^{2}}
\end{equation}

The position $y_c=y_T-a\cos\theta_T$ of the center of the disc is then expressed as
\begin{equation}
    \label{yc2D}
    y_c=\frac{a \left(2 \xi_T + 1\right)}{\left(\xi_T + 1\right)^{2}} - \frac{\sqrt{2} \lambda \xi_T}{\xi_T + 1},
\end{equation}
from which $h_0=y_c/a-1$ is calculated.

The normal force density $f_n$ applied by the interface on the disc in the contact region is composed of two parts:
\begin{equation}
\label{fn2D}
f_n(x)=\Delta\rho g[(\lambda^2/a+\sqrt{a^2-x^2}-y_c)],
\end{equation}
where the first term in the square brackets is the tension effect and the remaining terms are the gravity effect.
Using the relation $\boldsymbol n_i\boldsymbol\cdot\boldsymbol e^y dl_i/dx=1$, the total force is obtained.
The force applied by the disc on the interface is opposite to the force applied by the interface on the disc, which leads to the following expression for the downward force:
\begin{equation}
\label{Felastic2D}
    F_y=\Delta\rho g
        \left[
- a^{2} \operatorname{atan}{\left(\frac{\xi_T \sqrt{\xi_T^{2} + 4 \xi_T + 2}}{2 \xi_T + 1} \right)} + \frac{\xi_T \sqrt{\xi_T^{2} + 4 \xi_T + 2} \left(\frac{a^{2} \left(2 \xi_T + 1\right)}{\left(\xi_T + 1\right)^{2}} - \frac{2 \sqrt{2} a \lambda \xi_T}{\xi_T + 1} - 2 \lambda^{2}\right)}{\left(\xi_T + 1\right)^{2}}        
        \right]
\end{equation}

Taylor expansions of $h_0$ and $F_y$ for small values of $\xi_T$ are
\begin{equation}
    \label{Taylorh2D}
    h_0=- \sqrt{2} \lambda \xi_T/a +\xi_T^{2} \left(- 1 + \sqrt{2} \lambda/a\right)+ O\left(\xi_T^{3}\right),
\end{equation}
\begin{equation}
    \label{TaylorF2D}
F_y=- \Delta\rho g\left[2 \sqrt{2} \lambda^{2} \xi_T+\xi_T^{2} \left(- 4 a \lambda + 2 \sqrt{2} \lambda^{2}\right) + \xi_T^{3} \left(- \frac{4 \sqrt{2} a^{2}}{3} + 8 a \lambda - \frac{3 \sqrt{2} \lambda^{2}}{2}\right)  + O\left(\xi_T^{4}\right)\right],
\end{equation}
whence $F_y\approx 2\sigma h_0a/\lambda$ for $\lambda>0$ and $F_y/(\Delta\rho ga^2)\approx -2(-2h_0)^{3/2}/3$ for $\lambda=0$.

\subsection{Elastic indentation in 3D}
The setup of the 3D problem is the same as in the 2D scheme shown in Fig.\ref{fig:elastic} but now assuming rotational symmetry about the $y$ axis with $x$ referring to the distance from the symmetry axis in cylindrical coordinates.

For simplicity, the deformation problem is treated in the linear approximation in which the shape equilibrium equation for the free part of the interface is
\begin{equation}
\label{interface3D}
    \lambda^2\boldsymbol\nabla^2y_i-y_i=\lambda^2\frac{1}{x}\frac{d}{dx}\left(x\frac{dy_i(x)}{dx}\right)-y_i(x)=0,
\end{equation}
the solution to which and the boundary condition (\ref{bc2D1}) is
\begin{equation}
\label{y3D}
    y_i(x)=- bK_{0}\left(\frac{x}{\lambda}\right),
\end{equation}
where $b>0$ is an amplitude to be expressed below and $K$ refers to modified Bessel functions of the second kind.
The angle $\theta_T$ is used as the parametrization in the 3D case.
$x_T$ is calculated as
\begin{equation}
    \label{xT3D}
    x_T=a\sin\theta_T.
\end{equation}
The amplitude $b$ is found to be
\begin{equation}
    \label{A3D}
    b=\frac{\lambda \tan{\left(\theta_T \right)}}{K_{1}\left(\frac{a \sin{\left(\theta_T \right)}}{\lambda}\right)}
\end{equation}
$y_T$ is expressed by substituting $x=x_T$ into (\ref{y3D})
\begin{equation}
    \label{yT3D}
    y_T=- \frac{\lambda \tan{\left(\theta_T \right)} K_{0}\left(\frac{a \sin{\left(\theta_T \right)}}{\lambda}\right)}{K_{1}\left(\frac{a \sin{\left(\theta_T \right)}}{\lambda}\right)}
\end{equation}
and $y_c$ is expressed as $y_T+a\cos\theta_T:$ 
\begin{equation}
    \label{yc3D}
    y_c=a \cos{\left(\theta_T \right)} - \frac{\lambda \tan{\left(\theta_T \right)} K_{0}\left(\frac{a \sin{\left(\theta_T \right)}}{\lambda}\right)}{K_{1}\left(\frac{a \sin{\left(\theta_T \right)}}{\lambda}\right)}.
\end{equation}

The normal force expression is similar to the 2D case (\ref{fn2D}):
\begin{equation}
\label{fn3D}
f_n(x)=\Delta\rho g[(\lambda^2/a+\sqrt{a^2-x^2}-y_c)].
\end{equation}
Integrating the force expression over the contact region yields
\begin{equation}
\label{Felastic3D}
    F_y=-2 \pi\Delta\rho g\left(
        -\frac{a^{3} \sin^{2}{\left(\theta_T \right)} \cos{\left(\theta_T \right)}}{2}+
        \frac{a^{3}}{3}+
        \frac{a^{2}\lambda\sin^{2}{\left(\theta_T \right)} \tan{\left(\theta_T \right)}K_{0}\left(\frac{a \sin{\left(\theta_T \right)}}{\lambda}\right)}{2 K_{1}\left(\frac{a \sin{\left(\theta_T \right)}}{\lambda}\right)}+ 
        -\frac{a^{3} \cos^3{\left(\theta_T \right)}}{3} +
        a\lambda^{2}\sin^{2}{\left(\theta_T \right)}
    \right).
\end{equation}
Assuming $\lambda\gg a$ and $\theta_T\ll 1$, expression (\ref{Felastic3D}) simplifies to 
\begin{equation}
    \label{F3D}
    F_y\approx -2\pi\Delta\rho ga\lambda^2\theta_T^2=-2\pi\sigma a\theta_T^2.
\end{equation}
The indentation depth expression reduces to 
\begin{equation}
    \label{h3D}
    h_0=\left(\ln\frac{a\theta_T}{2\lambda}-\frac{1}{2}+\gamma\right)\theta_T^2,
\end{equation}
where $\gamma$ is Euler's constant.

\subsection{Reynolds equation in polar coordinates}
The lubrication approximation describes the flow in a fluid film whose thickness is much smaller than all other length scales of the problem.
This approximation is classically used to compute the lift force for a quasi-flat interface, assuming the deformation of the interface to be much smaller than the thickness of the film separating the particle and the interface.
In this section, the lubrication equation is formulated and analyzed for $h_0<0$.
Similarly to the elastic indentation problem, the polar coordinates $(r,\theta)$ centered about the particle center are used with the polar angle measured from the downward direction (as shown in Fig.3 of the main text).
The interface shape is parametrized by the film thickness function $H(\theta)$, which measures the length of the section of the film in the radial direction for a given polar angle $\theta$.
The flow in the film is measured in the reference frame co-moving with the particle in this derivation.
Only the tangential component $u_\theta$ of the flow needs to be considered to the leading order, for which only the terms up to quadratic in $r$ need to be retained:
\begin{equation}
    \label{lubrication_flow}
    u_\theta(r,\theta)=a\Omega+u_1(\theta)(r-a)+u_2(\theta)(r-a)^2
\end{equation}
The constant term in (\ref{lubrication_flow}) is equal to $a\Omega$, where $\Omega<0$ is the angular velocity of the particle due to the no-slip boundary condition at the particle boundary.
The boundary condition at the interface is more complicated and expresses the shear stress continuity between the fluid in the film and the fluid below the interface:
\begin{equation}
    \label{shear_stress}
    \partial_r u_\theta(\theta)=u_1(\theta)+2u_2(\theta)H(\theta)=s(\theta)/\eta,
\end{equation}
where $s(\theta)=O(\eta\dot\gamma)$ is the shear stress distribution at the interface.
In general, the explicit form of $s(\theta)$ is not known as it depends non-locally on the interface shape and the velocity at the interface.
However, numerical results suggest that $s$ scales as $O(\dot\gamma)$ for small $\dot\gamma$, and thus can be neglected compared to other terms.
Finally, the flux $q_0<0$ must be conserved inside the fluid film (i.e., it is independent of $\theta$):
\begin{equation}
    \label{flux_conservation}
    q_0=\int_a^{a+d}u_\theta(r)dr=a\Omega H+u_1(\theta)H^2/2+u_2(\theta)H^3/3.
\end{equation}
Solving eqs. (\ref{shear_stress}) and (\ref{flux_conservation}) together and neglecting $s$ yields an expression for $u_2$ as a function of two global constants $\Omega$ and $q_0$:
\begin{equation}
    \label{u2}
    u_2=\frac{3}{2}\frac{a\Omega H-q_0}{H^3}
\end{equation}
The $u_2$ component is of interest because it defines the pressure gradient within the film
\begin{equation}
    \label{pressure_gradient}
    2\eta u_2=\partial_\theta p(\theta)/a.
\end{equation}
The angular velocity $\Omega$ and the flux $q_0$ can be found using the zero-torque
\begin{equation}
    \label{zero_torque}
    \int_{\theta_i}^{\theta_o} u_1(\theta)d\theta=O(\dot\gamma)
\end{equation}
 and the zero-pressure-difference conditions:
\begin{equation}
    \label{zero_pressure_drop}
    \int_{\theta_i}^{\theta_o} \partial_\theta p d\theta=2\eta a\int_{\theta_i}^{\theta_o} u_2(\theta)d\theta=O(\eta\dot\gamma).
\end{equation}
Here $\theta_i$ and $\theta_o$ are the inlet and the outlet of the film, respectively.
Equating the pressure in the film to the elastic force generated by the deformed interface yields the following equation for the film thickness:
\begin{equation}
    \label{general_lubrication}
    \frac{\sigma}{a}\partial_\theta\left[\frac{(\partial_\theta H)^2+(a+H)^2-(a+H)\partial_{\theta\theta}H}{\sqrt{(\partial_\theta H)^2+(a+H)^2}^3}-B_o\left(h_0+a-(a+H)\cos\theta\right)\right]=
    3\frac{a\Omega H-q_0}{H^3}
\end{equation}

Retaining only the leading-order terms, eq. (\ref{general_lubrication}) simplifies to
\begin{equation}
    \label{indentation_lubrication}
    -\frac{\sigma}{a^3}\partial_{\theta\theta\theta}H=3\eta\frac{a\Omega H-q_0}{H^3}
\end{equation}
by assuming $H\ll a$, $\partial_\theta H\ll a$ and $H^2 s\ll q_0$ (as validated below).
Equation (\ref{indentation_lubrication}) has a trivial solution $H=H_*\equiv q_0/(a\Omega)$
Using this solution as the zero-order approximation, the linearized equation for $\delta H\equiv H-H_*$ is obtained as
\begin{equation}
    \label{indentation_linearized}
    -\frac{\sigma}{a^3}\partial_{\theta\theta\theta}\delta H=\frac{3\eta a\Omega}{H_*^3}\delta H
\end{equation}
The characteristic equation corresponding to eq. (\ref{indentation_linearized}) is $\kappa^3-\kappa_*^3=0$, where 
\begin{equation}
    \label{kappa}
    \kappa^3_*=-3\eta\Omega a^3/(H_*^3\sigma)
\end{equation}
is positive since $\Omega<0$.
The three roots of the characteristic equation correspond to $\kappa_*$ and $-\kappa_*/2\pm i\sqrt{3}\kappa_*/2$.

This suggests that the solution of the equation (\ref{indentation_lubrication}) corresponds to 3 sections:
In the central section, $\delta H\simeq 0$. Near the inlet, $\delta H\propto e^{\kappa_*\theta}$, which grows exponentially for increasing $\theta$.
Finally, the film thickness shows exponentially decreasing (as $\theta$ increases) oscillations in vicinity of the outlet.
The wave vector and the decrease rate scale as $\kappa_*$.
These oscillations can be recognized in the fluid film thickness graph in Fig. 3 of the main text.

The pressure in the central section must balance the surface tension force of the interface, which is of order $\sigma/a$ and is thus independent of $\dot\gamma$.
The pressure jump across the inlet section thus must be of the same order.
This leads to the following relation:
\begin{equation}
	\label{inlet}
	\sigma/a={3\eta a\Omega}\int\frac{H-H_*}{H^3}d\theta={3\eta a\Omega}\int_{-\infty}^\infty\frac{\delta H_0 e^{\kappa\theta}}{(H_*+\delta H_0 e^{\kappa\theta})^3}d\theta={3\eta a\Omega}\frac{1}{2H_*^2\kappa}={3\eta \Omega}\frac{1}{2H_*}\left(-\frac{\sigma}{3\eta\Omega}\right)^{1/3}
\end{equation}
for any amplitude $\delta H_0$,
whence
\begin{equation}
	\label{omega0_indent}
	\Omega=-\frac{\sigma}{3\eta}\left(\frac{2H_*}{a}\right)^{3/2}.
\end{equation}
In general, the approximation $H=H_*+\delta H_0 e^{\kappa\theta}$ is only valid near the inlet but in practice, eq. (\ref{omega0_indent}) shows not only the correct scaling for $\Omega$ in the limit $H_*\rightarrow 0^+$ but also a good approximation for the prefactor.

The analysis of the pressure drop in the outlet region is more difficult due to the oscillatory dependence of $H$ on $\theta$ near the outlet.
It is found, however, that these oscillations lead to some regions where $H<H_*$, which is necessary for the pressure to decrease as $\theta$ decreases (going across the outlet from the middle section of the film).
It is further found that the oscillations of $H$ grow exponentially as $\theta$ decreases only up to a certain amplitude, at which point the non-linear effects take over leading to a steady growth of $H$ with decreasing $\theta$.
The pressure jump across the outlet region is thus mainly determined by the last (counting in decreasing $\theta$ direction) region of $H<H_*$.
The length of this region is of order $a/\kappa_*$ (wave-length of the oscillations of $H$ in the linear regime) and the amplitude of the last oscillation scales as $H_*$.
It is thus possible to see that the pressure jump across the outlet region is of the same order $a\Omega\kappa/H_*^2$ as across the inlet region but has the opposite sign.
This explains how the pressure in the fluid film between the particle and the interface can be sufficiently high to maintain the particle at a constant separation from the indented interface even at weakest flow.

\begin{figure}
    \begin{center}
        \includegraphics[width=0.45\columnwidth]{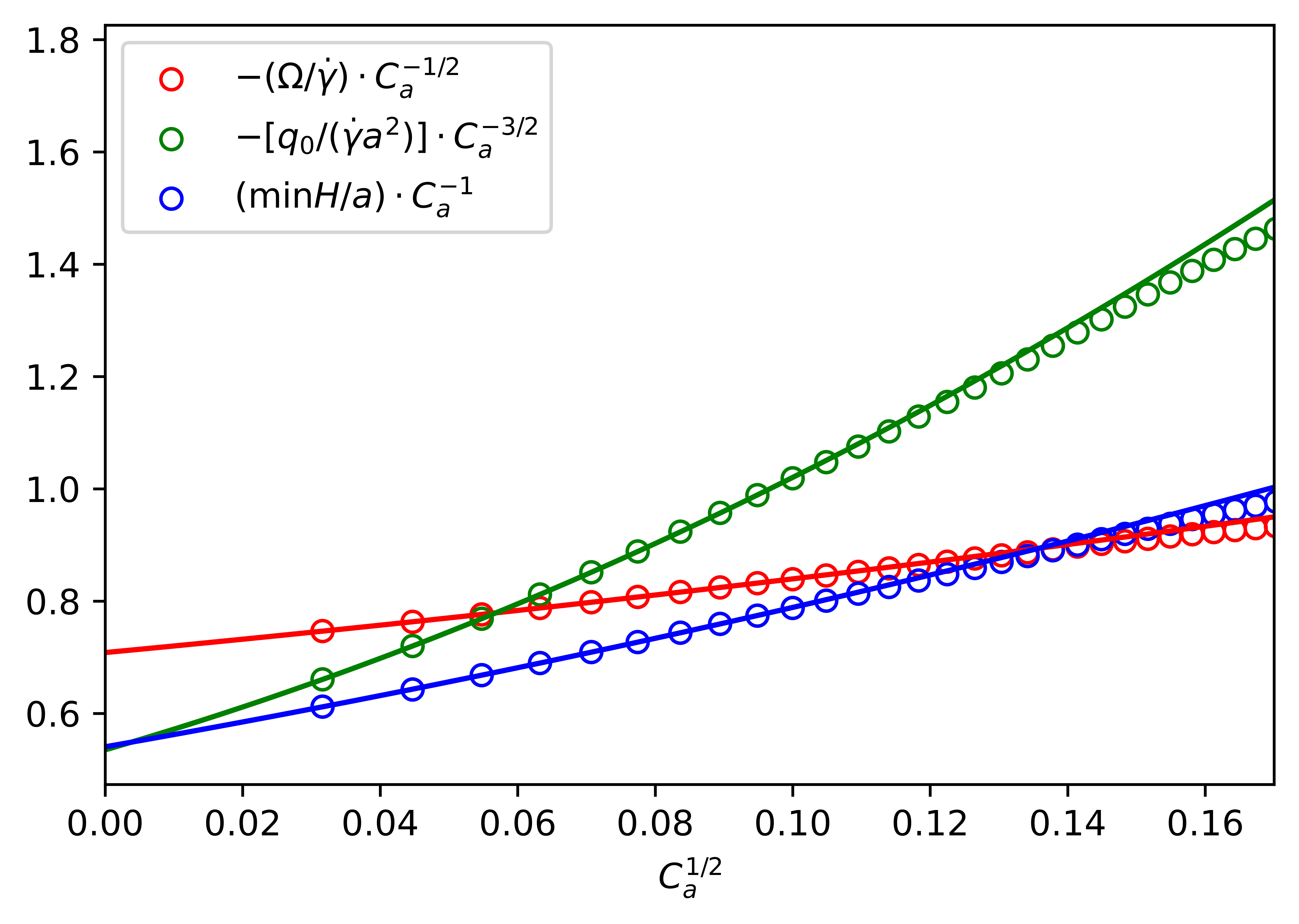}
        \caption{\label{fig:polar_scaling} Comparison of theoretical scaling laws for $\min{H}/aC_a^{-1}\propto\min{H}/\dot\gamma$, $\Omega/\dot\gamma C_a^{-1/2}\propto \Omega/\dot\gamma^{3/2}$, and $q_0/(a^2\dot\gamma)C_a^{-3/2}\propto q_0/\dot\gamma^{5/2}$ with $\dot\gamma$ with numerical results. Symbols are numerical results, solid curves are second-degree polynomial fits. The plotted curves are expected to have a finite limit for $C_a\rightarrow 0^+$.
        }
    \end{center}
\end{figure}

Finally, the scaling of $H_*$ as a function of $\dot\gamma$ can be found by computing the torque acting on the particle:
Combining eq. (\ref{shear_stress}) (neglecting $s$) and eq. (\ref{u2}) shows that
\begin{equation}
    \label{u1_indention}
    u_1(\theta)=-6a\Omega\frac{H-H_*}{H^2}
\end{equation}
Similarly to the pressure distribution, the main contribution to the torque comes from the inlet and the outlet regions.
Unlike pressure jumps, however, these two contributions do not balance each other:
They have the same scaling for small $H_*$ but different prefactors.
The scaling of the shear stress integral near the inlet can be estimated by substituting $H=H_*+\delta H_0\exp\kappa\theta,$ similarly to the pressure jump calculation.
Combining this estimate with eq. (\ref{zero_torque}) leads to the following relation
\begin{equation}
    \label{zero_torque_indentation}
    a\Omega\eta/(\kappa H_*)=O(\eta\dot\gamma),
\end{equation}
whence by comparison with second-from-right equality in eq. (\ref{inlet}),
\begin{equation}
    \label{dstar}
    H_*\propto\eta\dot\gamma/\sigma
\end{equation}
It follows thus that for fixed indentation depth, the film thickness scales as $O(\dot\gamma)$, the length of the inlet and the outlet regions scale as $\kappa_*^{-1}=O(\dot\gamma^{1/2})$, the angular velocity scales as $O(\dot\gamma^{3/2})$ and the flux in the fluid film scales as $O(\dot\gamma^{5/2})$.
This justifies neglecting the shear stress $s$ in eq. (\ref{shear_stress}).
The derived scaling laws are validated by numerical results in Fig.\ref{fig:polar_scaling}.

\subsection{Exact solution for a flat interface}
\subsubsection{General discussion}
Unlike for the strongly-indented interface, the stress in the lower fluid $s$ can not be neglected for a quasi-flat interface.
This stress, however, can not be calculated in the lubrication approximation because it varies on the length scale of the order of the particle size.
Furthermore, the angular velocity of the particle needs to be computed from the balance of torques acting on the particle, which includes non-negligible contributions from flows outside of the thin film of fluid separating the particle and the interface.
It is shown here that (1) the shear stress at the interface can be written as $s(x)=2\eta\dot\gamma\left[1+O(h_0^{1/2})\right]$ for $x\ll a$ and (2) the angular velocity of the particle can be written as $\Omega=-4\dot\gamma h_0\left[1+O(h_0^{1/2})\right]$ in the flat-interface limit.
These non-trivial results are obtained using the exact solution of the problem in bipolar coordinates with flow fields represented by analytic functions.

The deformation of the interface is negligible in the leading-order approximation and the surface-tension/gravity forces act as Lagrange multipliers that ensure the consistency of the hydrodynamic problem.
In practice, this means that only the tangential component of the stress is continuous across the interface.

Throughout this section 2D vectors are represented by complex numbers, identifying $x$ and $y$ components as real and imaginary parts, respectively.
In particular, $\zeta=x+iy$ represents the Cartesian coordinates.
For simplicity, the solution is presented for $a=1$ only.
It is generally known that solutions of the Stokes equations in a given domain can be parameterized by two functions $M$ and $N$ of complex variables, analytic on the respective domain:
\begin{equation}
    \label{complex_parametrization}
    u(\zeta)=M(\zeta)-\zeta\overline{M'(\zeta)}+\overline{N(\zeta)}.
\end{equation}
Here bars refer to complex conjugation and prime refers to differentiation with respect to the complex variable.
The bipolar coordinates are set up by parameterizing $\zeta$ as
\begin{equation}
    \label{conformal}
    \zeta=i\frac{\rho_0^2-1}{2\rho_0}\frac{w-1}{w+1},
\end{equation}
where $w$ is another complex variable and $\rho_0$ is a parameter that is used below to set the distance between the particle and the interface.
The main advantage of the map (\ref{conformal}) is that it maps the circle $|w|=1$ to the interface ($\Im z=0$) and the circle $|w|=\rho_0$ to a circle of radius 1 centered at $\zeta_0=i(\rho_0+1/\rho_0)/2$.
The non-dimensional gap is then equal to $\Im \zeta_0-1=(\rho_0+1/\rho_0)/2-1$.
The pullback of the parameterization (\ref{complex_parametrization}) on the $w$ domain is written as
\begin{equation}
    \label{complex_parametrization2}
    u(w)=\tilde M(w)-\frac{\zeta(w)}{\overline{\zeta'(w)}}\overline{\tilde M'(w)}+\overline{\tilde N(w)}+\frac{\overline w^2-1}{2}\overline{\tilde M'(w)},
\end{equation}
where $\tilde M$ and $\tilde N$ are two complex functions of $w$ analytic for $w<1$ if $u$ is the flow in the lower fluid or in the annulus $1<w<\rho_0$ if $u$ is the flow in the upper fluid.
The last term in eq. (\ref{complex_parametrization2}), which is an analytic function of $\overline w$ on the same domain as $\overline{\tilde N(w)}$ is split off for convenience to rewrite the velocity expression as
\begin{equation}
    \label{complex_parametrization3}
    u(w)=\tilde M(w)-\Phi(w,\overline w)\overline{\tilde M'(w)}+\overline{\tilde N(w)},
\end{equation}
where
\begin{equation}
    \label{shape}
    \Phi=\frac{\zeta(w)}{\overline{\zeta'(w)}}-\frac{\overline w^2-1}{2}
\end{equation}
possesses a convenient property $\Phi=0$ for $|w|=1$.
This greatly simplifies imposing the boundary conditions at the fluid-fluid interface.

The flows in the lower and upper fluids are written as $u=u^\infty+u_2$ and $u=u^\infty+u_1$, respectively, where
\begin{equation}
    \label{uinfinity}
    u^\infty=\dot\gamma i(\overline \zeta-\zeta)/2
\end{equation}
is the imposed shear flow.
The velocities $u_1$ and $u_2$ are parameterized as defined in eq. (\ref{complex_parametrization3}) with $\tilde M$ and $\tilde N$ denoted as $M_1$ and $N_1$ or $M_2$ and $N_2$, respectively.

The variable $w$ is decomposed as $w=\rho e^{i\phi}$ to apply the boundary conditions at $\rho=1$ (fluid-fluid interface) and $\rho=\rho_0$ (particle boundary).

The velocities $u_1$ and $u_2$ at the interface are real (along the $x$ direction) and are invariant under the  $x\rightarrow-x$ transformation.
They can thus be written as
\begin{equation}
    \label{u1}
    u_{1,2}(e^{i\phi})=u_0+\sum_{k=1}^\infty U_k\cos(k\phi)
\end{equation}
with unknown amplitudes $U_k$.
The functions $M_2$ and $N_2$ are analytic for $\rho<1$ and thus can be expanded in Taylor series of $w$.
It is straightforward to show that 
\begin{equation}
    \label{M2N2}
    M_2(w)=N_2(w)=\frac{u_0}{2}+\frac{1}{2}\sum_{k=1}^\infty U_kw^k
\end{equation}
from the boundary condition (\ref{u1}).

The functions $M_1$ and $N_1$ are only analytic in an annular region and thus must be represented by Laurent series:
\begin{equation}
    \label{M1N1}
    M_1(w)=C_0+\sum_{k=1}^\infty (C_kw^k+E_kw^{-k}),\,\,\,N_1(w)=\sum_{k=1}^\infty (D_kw^k+B_kw^{-k}).
\end{equation}
Note that there could be terms proportional to $\ln w$ in the $M_1(w)$ and $N_1(w)$ expansions but they are absent for a force-free particle.

Besides the boundary condition (\ref{u1}), there is a continuity of tangential stress across the interface.
The mappings of the complex plane by analytic functions, such as the one given by eq. (\ref{conformal}), are known to be conformal, and as such, map normals to normals and tangent vectors to tangent vectors.
The continuity of the tangential stress is thus written as
\begin{equation}
    \label{tangential_stress}
    \partial_\rho u_1(\rho,t)=\partial_\rho u_2(\rho,t)\textrm{ for }\rho=1.
\end{equation}
Applying the two conditions (\ref{u1}) and (\ref{tangential_stress}) gives 3 conditions for $U_k$, $C_k$, $D_k$, $E_k$ and $B_k$ for each $k>0$.
It is thus possible to express $U_k$, $D_k$, and $B_k$ as
\begin{equation}
    \label{UDFk}
    U_k=2(C_k-E_k),\,\,\,D_k=C_k-2E_k,\,\,\,B_k=-E_k,\textrm{ for }k>0.
\end{equation}

The values of $C_k$ and $E_k$ should then be found by imposing the rigid-body motion at the particle boundary:
\begin{equation}
    \label{u2particle}
    u_1(\rho_0,\phi)+u^\infty(\rho_0,\phi)=V_0+i\Omega[\zeta(\rho_0,\phi)-\zeta_0]
\end{equation}
This is the most difficult part of the problem since the function $\Phi$ is not equal to 0 at the particle boundary.
As a consequence, the equations for different $k$ are entangled and only the special simplicity of the problem allows it to be solved exactly, as detailed below.

The function $\Phi$ can be expanded in harmonics of $\phi$ for $\rho>1$ as follows
\begin{equation}
    \label{Fexp}
    \Phi(\rho,\phi)=\left(\rho^{-1}-\rho\right)e^{-i\phi}-(\rho^2-1)^2\sum_{k=2}^\infty e^{-ik\phi}\rho^{-k}.
\end{equation}
Note that only negative powers of $e^{i\phi}$ enter (\ref{Fexp}).

The system of equations for a given $k>0$ is then written as
\begin{equation}
    \label{CEeq1}
    C_k\rho_0^k+E_k\left[k\rho_0^{-k-2}-(k+1)\rho_0^{-k}\right]+\left(\rho_0-\rho_0^{-1}\right)^2\rho_0^k\sum\limits_{l=k+1}^\infty lE_l\left(-\rho_0^2\right)^l=\dot\gamma(-\rho_0)^{-k}\frac{\rho_0^2-1}{2\rho_0}
\end{equation}
for the $k$-th harmonic of eq. (\ref{u2particle}) and
\begin{equation}
    \label{CEeq2}
    C_k\left[(k+1)\rho_0^k-k\rho_0^{k-2}\right]+E_k\left(\rho_0^{-k}-2\rho_0^k\right)+\left(\rho_0-\rho_0^{-1}\right)^2\rho_0^k\left[\sum\limits_{l=1}^\infty lE_l\left(-\rho_0^2\right)^l-\sum\limits_{l=1}^{k-1}lC_l(-\rho_0^2)^l\right]=(-1)^k\left(2\Omega+\dot\gamma\right)\frac{\rho_0^2-1}{2\rho_0^{k+1}}
\end{equation}
for the $-k$-th harmonic.
As can be seen, eqs. (\ref{CEeq1}) and (\ref{CEeq2}) are both forward-and backward-entangled, which means that some form of truncation needs to be used to solve them.
However, there is a special way to eliminate the forward entanglement:
The trick is to use the sum
\begin{equation}
    \label{Sinf}
    S_\infty=\sum\limits_{l=1}^\infty lE_l\left(-\rho_0^2\right)^l
\end{equation}
as an additional unknown in the problem, rewriting eqs. (\ref{CEeq1}) and (\ref{CEeq2}) as
\begin{equation}
    \label{CEeq3}
    C_k\rho_0^k+E_k\left[k\rho_0^{-k-2}-(k+1)\rho_0^{-k}\right]+\left(\rho_0-\rho_0^{-1}\right)^2\rho_0^k\left[S_\infty-\sum\limits_{l=1}^k lE_l\left(-\rho_0^2\right)^l\right]=\dot\gamma(-\rho_0)^{-k}\frac{\rho_0^2-1}{2\rho_0}
\end{equation}
and
\begin{equation}
    \label{CEeq4}
    C_k\left[(k+1)\rho_0^k-k\rho_0^{k-2}\right]+E_k\left(\rho_0^{-k}-2\rho_0^k\right)+\left(\rho_0-\rho_0^{-1}\right)^2\rho_0^k\left[S_\infty-\sum\limits_{l=1}^{k-1}lC_l(-\rho_0^2)^l\right]=(-1)^k\left(2\Omega+\dot\gamma\right)\frac{\rho_0^2-1}{2\rho_0^{k+1}},
\end{equation}
which is only backward-entangled (equations for a given $k$ depend only on $C_l$ and $E_l$ with $l\le k$, $S_\infty$ and $\Omega$).
It turns out that the system of eqs. (\ref{CEeq3}) and (\ref{CEeq4}) at $k=1$ is defective in terms of unknowns $C_1$ and $E_1$.
Instead, it can be solved for $S_\infty$ and $C_1$.
The higher-order systems can be unraveled one order at a time using the solutions for the lower orders to compute all $E_k$s with $k>1$ and all $C_k$s as a function of $\Omega$ and $E_1$.
The value of $E_1$ can then be found from the consistency of eq. (\ref{Sinf}) and $\Omega$ can be found from the zero-torque condition for the particle.
In practice, it is possible to avoid finding $E_1$ explicitly when calculating $\Omega$ or the shear stress at the interface.

\subsubsection{Calculating the angular velocity}
\begin{figure}
    \begin{center}
        \includegraphics[width=0.45\textwidth]{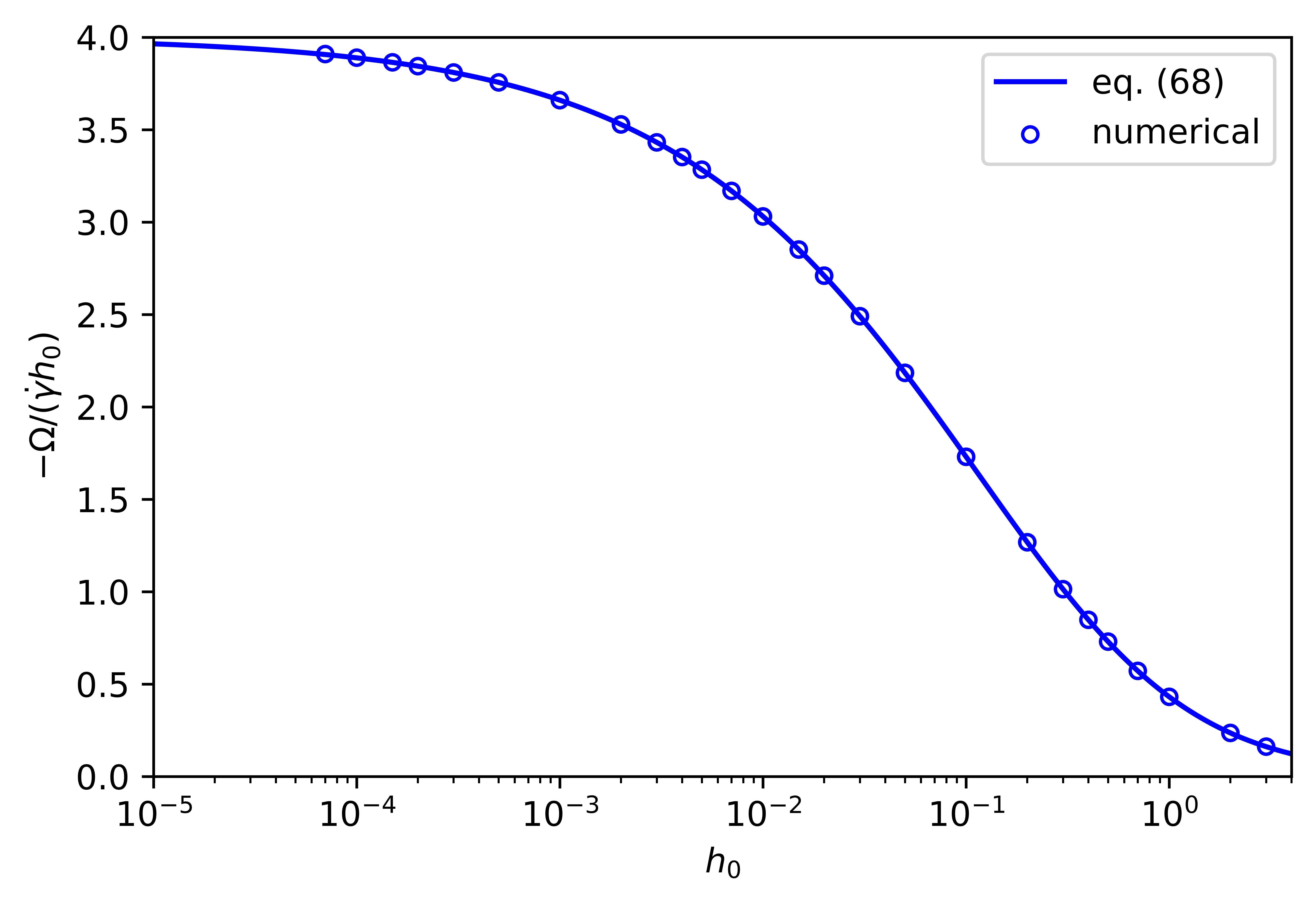}
        \includegraphics[width=0.45\textwidth]{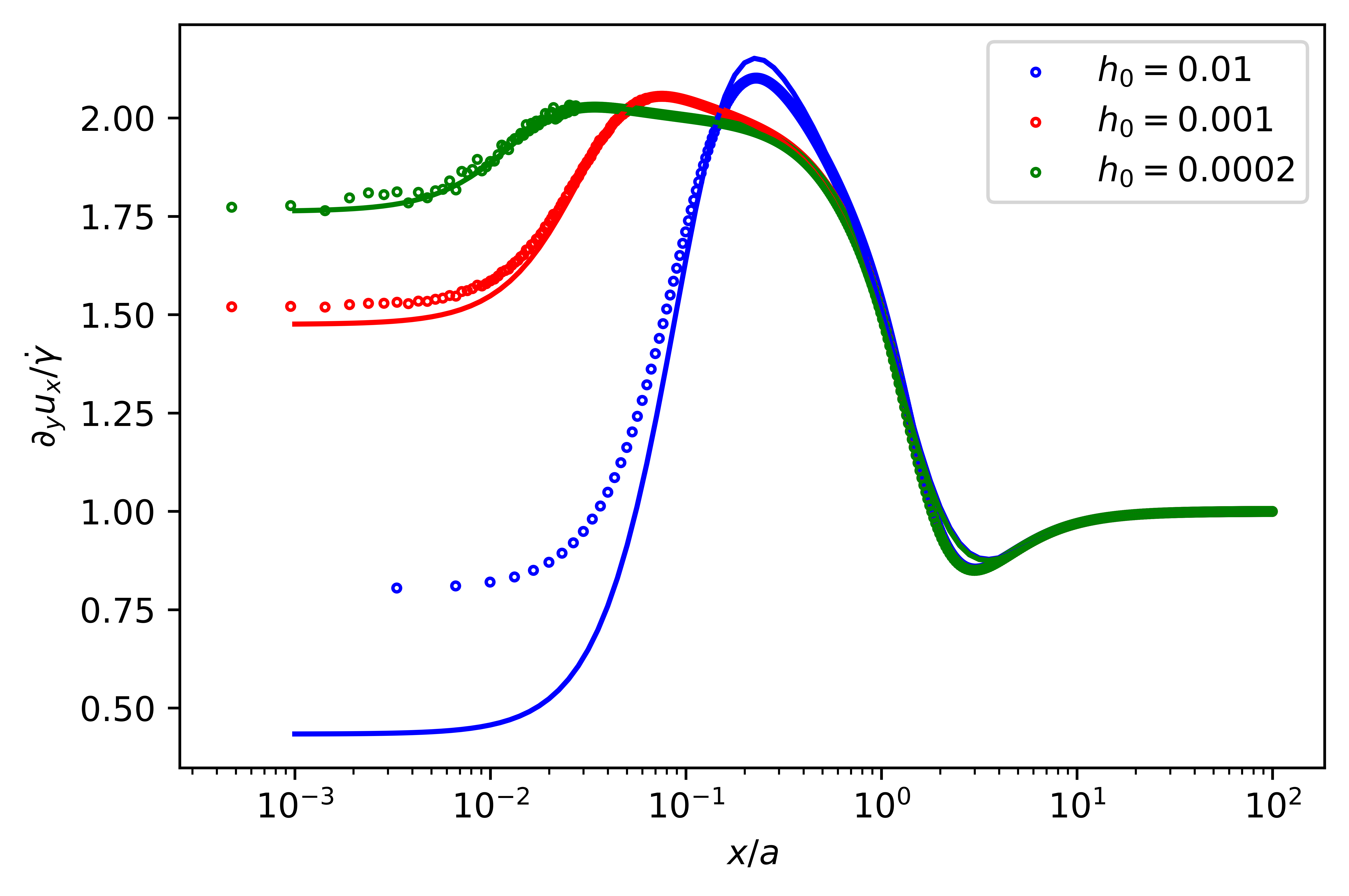}
        \caption{\label{fig:exact}Comparison of the exact solution of the hydrodynamic problem for a flat interface with numerical results. Left: Angular velocity $\Omega$. Right: Shear stress distribution at the interface. Symbols are numerical results, solid lines are eq. (\ref{stress2}). Note that the systematic discrepancy for $x/a>1$ is due to neglecting the higher-order terms in eq. (\ref{uk}).}
    \end{center}
\end{figure}

The torque (per unit length in 3rd dimension) $T_0$ applied on the fluid by the particle is computed as
\begin{equation}
    \label{torque0}
    T_0=-\frac{\rho(\rho_0^2-1)}{\rho_0(\rho^2-1)}\int_0^{2\pi} \frac{\Im (\partial_\rho u \overline{\partial_\phi \zeta}+\partial_\phi u \overline{\partial_\rho \zeta})}{|\partial_\rho \zeta|}d\phi=
    4\pi\frac{\rho_0^2-1}{\rho_0}\sum\limits_{k=1}^\infty (-1)^kk(C_k-D_k),
\end{equation}
where the integral is the same for any fixed $\rho\in(1,\rho_0)$.
Hence according to eq. (\ref{UDFk}),
\begin{equation}
    \label{torque}
    T_0=8\pi\frac{\rho_0^2-1}{\rho_0}\sum\limits_{k=1}^\infty (-1)^kkE_k=4\pi\frac{\rho_0^2-1}{\rho_0}\sum\limits_{k=1}^\infty (-1)^kk(k+1)(E_k+E_{k+1}),
\end{equation}
where the last equality in (\ref{torque}) is used because the sums $E_k+E_{k+1}$ can be evaluated without knowing $E_1$ (are independent of $E_1$) for all $k\ge 1$.
Furthermore, $E_k+E_{k+1}$ is independent of $\Omega$ for all $k>1$.
Finally, it is found that setting $\Omega=0$ in eq. (\ref{torque}) and expanding the result in powers of $1/\rho_0$ leads to almost all powers of $1/\rho_0$ disappearing from the expansion.
As a consequence, eq. (\ref{torque}) reduces to a very simple form $T_0=4\pi\rho_0^2\Omega/(\rho_0^2-1)+2\pi\dot\gamma(1-1/\rho_0^2),$ from which $\Omega$ is found as
\begin{equation}
    \label{Omega}
    \Omega=-\frac{(\rho_0^2-1)^2\dot\gamma}{2\rho_0^4}
\end{equation}
Comparing eq. (\ref{Omega}) with numerical results (Fig.\ref{fig:exact}, left) shows a very good agreement for all tested $h_0$, further corroborating the exactness of the solution (\ref{Omega}).

Expanding for $\rho_0$ close to 1, $\Omega=-2\dot\gamma(\rho_0-1)^2+O[(\rho_0-1)^3]=-4\dot\gamma h_0+O(h_0^{3/2})$.

\subsubsection{Calculating the shear stress at the interface}

The shear stress at the interface can be shown to be written as
\begin{equation}
    \label{stress}
    \eta\partial_y u_x(\phi)=\eta\dot\gamma+\left.\frac{\partial_\rho u_2(\rho,\phi)|}{|\partial_w \zeta|}\right|_{\rho=1}=\eta\dot\gamma+\frac{2\eta \rho_0(1+\cos \phi)}{\rho_0^2-1}\sum\limits_{k=1}^\infty 2k U_k\cos(k\phi).
\end{equation}
It must first be noted that $U_k$, is conveniently independent of $E_1$ for all $k$, which means that $U_k$ is given by an explicit function of $\rho_0$ for all $k>0$.
The remaining challenge consists in calculating the sum in eq. (\ref{stress}) in the limit $\rho_0\rightarrow 1^+$.

There are two strategies for evaluating the infinite sum in eq. (\ref{stress}) for a given $\rho_0$: First, only a finite number of terms can be taken, which provides a good approximation for large $\rho_0$ since $U_k$ scales roughly as $\rho_0^{-2k}$.
The second option is to expand each $U_k\rho_0^k$ into a power series of $\rho_0-1$. It turns out that the coefficients of such expansion can be represented as a polynomial of $k$ (for $k>1$) of degree that does not exceed the corresponding degree of $\rho_0-1$.
Such infinite sums can then be computed exactly.
This approach can be used to approximate the solution for $\rho_0$ close to 1, which is the case relevant for this study.

It is found that
\begin{equation}
    \label{uk}
    U_k=-\rho_0^{-k}\dot\gamma\begin{cases}
        -\Omega/\dot\gamma-(\rho_0-1)+3(\rho_0-1)^2/2-2(\rho_0-1)^3+...&\textrm{ for }k=1\\
        (-1)^k\left[(\rho_0-1)-(\rho_0-1)^2/2+(\rho_0-1)^3/2+...\right]&\textrm{ for }k>1.
    \end{cases}
\end{equation}
Note that the $(\rho_0-1)^4$ terms lose the simplicity of eq. (\ref{uk}) and are given by a cubic polynomial of $k$ multiplied by $(-\rho_0)^{-k}$.

Substituting eq. (\ref{uk}) into eq. (\ref{stress}) and taking the summation yields
\begin{equation}
    \label{stress2}
    \begin{aligned}
        \partial_y u_x(\phi)&=\dot\gamma-\frac{4\rho_0(1+\cos \phi)\dot\gamma}{\rho_0^2-1}\left[-\frac{\Omega}{\dot\gamma}+(\rho_0-1)^2-\frac{3(\rho_0-1)^3}{2}\right]\cos\phi+\\
        &\frac{2\dot\gamma\rho_0(1+\cos \phi)}{\rho_0^2-1}\left[(\rho_0-1)-\frac{(\rho_0-1)^2}{2}+\frac{(\rho_0-1)^3}{2}\right]\frac{2\rho_0+(\rho_0^2+1)\cos \phi}{\left(1+\rho_0^2+2\rho_0\cos \phi\right)^2}.
    \end{aligned}
\end{equation}
Since $\Omega=O[(\rho_0-1)^2]$, as shown above, it does not contribute to the leading-order approximation of $\partial_y u_x$ at the interface.
With this observation, the leading order approximation is
\begin{equation}
    \label{stress_simple}
    \partial_y u_x(\phi)=\dot\gamma+2\dot\gamma(1+\cos \phi)\frac{(1+\rho_0^2)\cos \phi +2\rho_0}{(1+\rho_0^2+2\rho_0\cos \phi)^2}.
\end{equation}
Substituting $\phi=2\tan^{-1}[2\rho_0x/(\rho_0^2-1)]$ and taking the limit $\rho_0\rightarrow 1^+$, a very simple expression
\begin{equation}
    \label{stress_simple2}
    \partial_y u_x(x)=\dot\gamma+\dot\gamma\frac{4a^2(4a^2-x^2)}{(4a^2+x^2)^2}
\end{equation}
is obtained.
This shows that $\partial_y u_x$ at the interface is equal to $2\dot\gamma$ for $x\ll a$ and $h_0\ll 1$.

Note that the denominator of eq. (\ref{stress2}) becomes small as $\phi$ approaches $\pm\pi$.
This problem becomes even worse for higher-order terms in $\rho_0-1$ in eq. (\ref{uk}), neglected in eq. (\ref{stress2}).
Since these terms are not constant but polynomial in $k$, they result in higher powers of $1+\rho_0^2+2\rho_0\cos \phi$ in the denominator, when summed together.
As a consequence, all terms in the formal expansion (\ref{stress2}) become comparable for $1-\cos \phi\gtrsim \rho_0-1$, which corresponds to $|x|/a\gtrsim 1$ regardless of $\rho_0$.
This problem is responsible for the small but noticeable discrepancy between the numerical results and equation (\ref{stress2}) in Fig.\ref{fig:exact}.
Since the shear stress at the interface is only important near the lubrication region for the purposes this study, this limitation of eq. (\ref{stress2}) is not critical.

\subsection{Classical lubrication approximation}
This section derives the $O(\dot\gamma^2)$ term of the downward force in the limit $h_0\rightarrow 0^+$.
The derivation follows the procedure outlined above for the indented interface, except that it is more convenient to work in the Cartesian coordinates.

\begin{figure}
    \begin{center}
        \includegraphics[width=0.45\columnwidth]{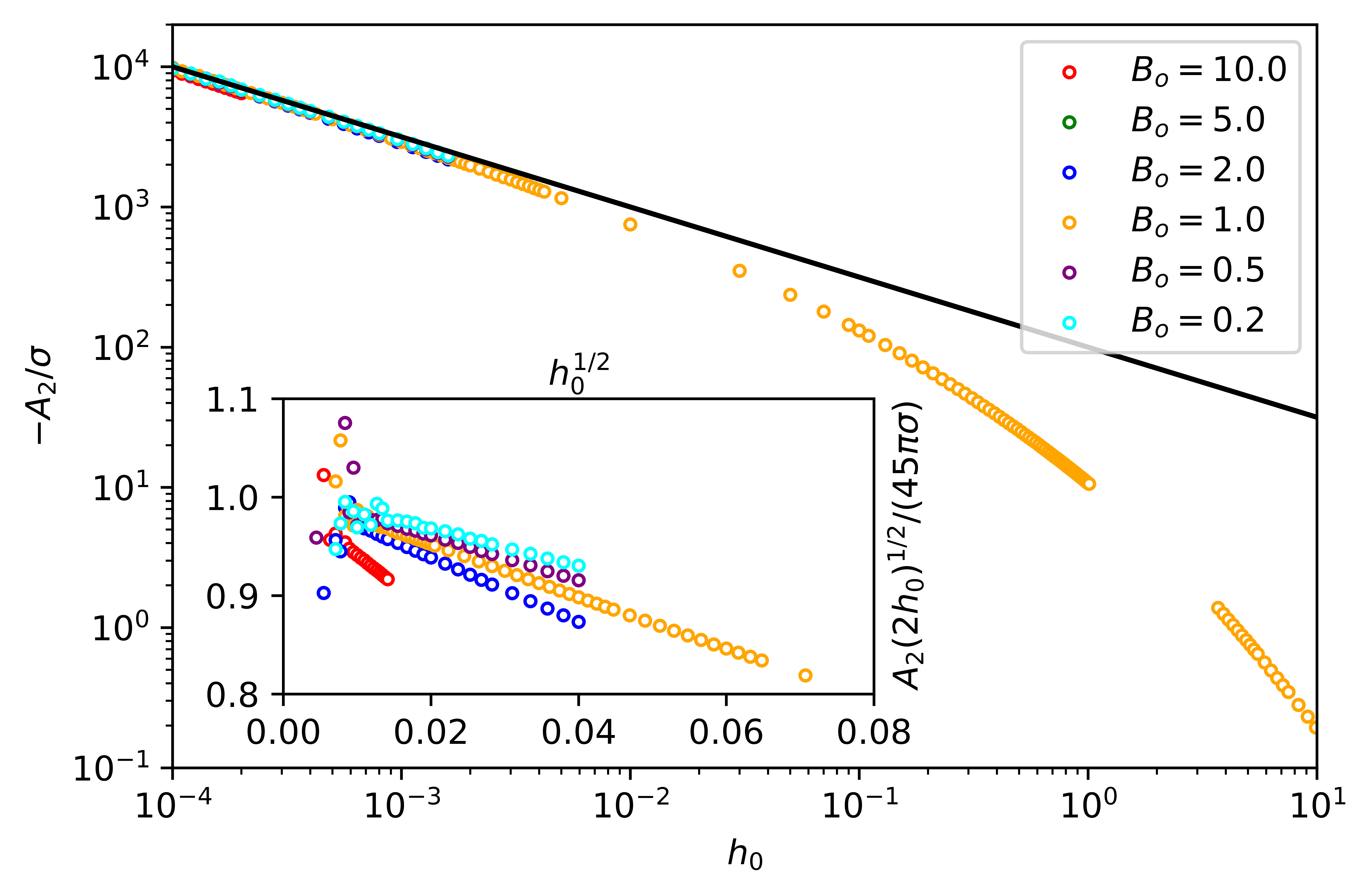}
        \includegraphics[width=0.45\columnwidth]{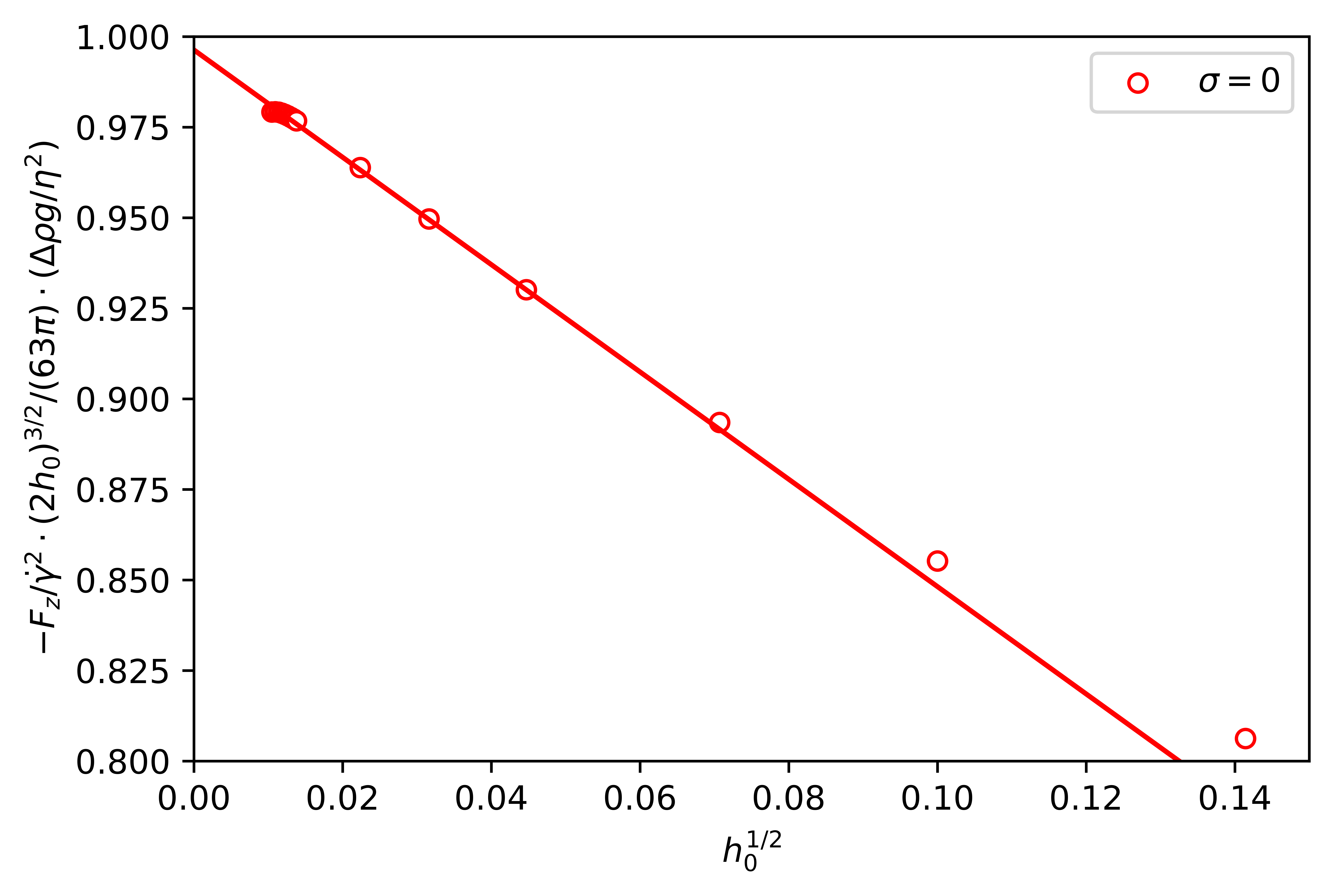}
        \caption{\label{fig:A2scaling} Comparison of eqs. (\ref{classic_Fourier_tension}) and (\ref{classic_Fourier_gravity}) with numerical simulations. Left: $A_2$ as a function of $h_0$ for finite $B_o$. Symbols are numerical results, solid line is eq. (\ref{classic_Fourier_tension}). Inset: Numerically computed ratio of the left and right hand sides of eq. (\ref{classic_Fourier_tension}) as a function of $h_0^{1/2}$ (theoretical limit for $h\rightarrow 0^+$ should be 1). Same legend as in the main plot. Right: Numerically computed ratio of the left and right hand sides of eq. (\ref{classic_Fourier_gravity}) as a function of $h_0^{1/2}$ (theoretical limit for $h\rightarrow 0^+$ should be 1). Some uncertainty in the numerical results is expected for $h_0$ around $10^{-4}$.
        }
    \end{center}
\end{figure}

As is classically done in the lubrication calculations, the particle boundary is approximated as $y_s(x)=ah_0+x^2/(2a)$.
The fluid film thickness is then written as $H(x)=y_s(x)-y_i(x)$.
The velocity field in the fluid film is written according to the classical lubrication approach as $\boldsymbol u(x,y)=\boldsymbol e^x[u_0(x)+u_1(x)y+u_2(x)y^2]$ in the reference frame co-moving with the particle.
The parameters $u_0$ $u_1$ and $u_2$ are calculated from the following constraints:
\begin{equation}
    \label{classic_bc}
    u_x(H)=-4\dot\gamma h_0,\,\,\,\partial_y u_x(0)=2\dot\gamma,\,\,\,\int_{0}^{H} u_xdy=q_0,
\end{equation}
where the first two constraints are extracted from the exact solution above and the last one is the conservation of the flux in the film.
Solving eqs. (\ref{classic_bc}) yields
\begin{equation}
    \label{classic_flow}
    u_0=2 \dot\gamma a h_{0} - \frac{\dot\gamma H}{2} + \frac{3 q_{0}}{2 H},\,\,\,u_1=2\dot\gamma,\,\,\,u_2=- \frac{6 \dot\gamma a h_{0}}{H^{2}} - \frac{3 \dot\gamma}{2 H} - \frac{3 q_{0}}{2 H^{3}},
\end{equation}
where $2\eta u_2=\partial_x p$ defines the pressure distribution inside the fluid film.
The value of $q_0$ is found by imposing zero-pressure-difference between the inlet and the outlet of the film:
\begin{equation}
    \label{classic_q0}
    q_0=-8\dot\gamma a^2h_0^2,
\end{equation}
whence the pressure gradient is written as
\begin{equation}
    \label{classic_dp}
    \partial_xp=\eta\dot\gamma\left[\frac{24a^2h_0^2}{(y_s-y_i)^3}-\frac{12ah_0}{(y_s-y_i)^2}-\frac{3}{y_s-y_i}\right].
\end{equation}
The pressure $p$ determines the force applied by the fluid on the interface in lubrication approximation.
Since the leading order of the problem corresponds to a pressure field that is anti-symmetric under $x\rightarrow -x$ transformation, the lift force is zero for a flat interface.
Consequently, the next-order term needs to be calculated in the expansion of eq. (\ref{classic_dp}) in powers of $y_i/y_s$:
\begin{equation}
    \label{classic_dp2}
    \partial_xp=\eta\dot\gamma\left[\frac{24a^2h_0^2}{y_s^3}-\frac{12ah_0}{y_s^2}-\frac{3}{y_s}\right]+
    \eta\dot\gamma y_i\left[\frac{72a^2h_0^2}{y_s^4}-\frac{24ah_0}{y_s^3}-\frac{3}{y_s^2}\right]+O(\dot\gamma^3)=\partial_x p_0-\frac{ay_i}{x}\partial_{xx} p_0+O(\dot\gamma^3),
\end{equation}
where $p_0$ is the solution of eq. (\ref{classic_dp}) for undeformed interface ($y_i=0$).
Using the second equality in (\ref{classic_dp2}), the pressure integral can be computed as
\begin{equation}
    \label{intp_classic}
    F_y=-\int_{-\infty}^\infty pdx=\int_{-\infty}^\infty x\partial_x pdx=-\int_{-\infty}^\infty ay_i\partial_{xx} p_0+O(\dot\gamma^3)
\end{equation}

The last integral in (\ref{intp_classic}) can be computed in Fourier space according to the Plancherel formula as
\begin{equation}
    \label{classic_Fourier}
    F_y=\frac{a}{\pi}\int_0^\infty k^2y_i(k)p_0(k) dk,
\end{equation}
where
\begin{equation}
    \label{yk}
    y_i(k)=\int_{-\infty}^\infty y_i(x)\sin(kx)dx
\end{equation}
and
\begin{equation}
    \label{pk}
    p_0(k)=\int_{-\infty}^\infty p_0(x)\sin(kx)dx=\frac{1}{k}\int_{-\infty}^\infty \partial_x p_0(x)\cos(kx)dx=6\pi\dot\gamma a\left[1+(2h_0)^{1/2}ak\right]e^{-ak\sqrt{2h_0}}.
\end{equation}
The gravity effect can be neglected for $\sigma>0$, which leaves $y_i(k)=-p_0(k)/(k^2\sigma)$:
\begin{equation}
    \label{classic_Fourier_tension}
    F_y=-\frac{a}{\pi}\int_0^\infty \frac{p_0(k)^2}{\sigma} dk=-\frac{45\pi\dot\gamma^2\eta^2a^2}{\sigma(2h_0)^{1/2}}=-\sigma\frac{45\pi C_a^2}{(2h_0)^{1/2}},
\end{equation}
For $\sigma=0$, the interface height is given by $y_i(k)=-p_0(k)/(\Delta\rho g)$ and the resulting force is
\begin{equation}
    \label{classic_Fourier_gravity}
    F_y=-\frac{a}{\pi}\int_0^\infty \frac{p_0(k)^2}{\Delta\rho g}k^2 dk=-\frac{63\pi\dot\gamma^2\eta^2a^2}{\Delta\rho g(2h_0)^{3/2}},
\end{equation}

\subsection{Zero-tension limit}

\begin{figure}
    \begin{center}
        \includegraphics[width=0.45\columnwidth]{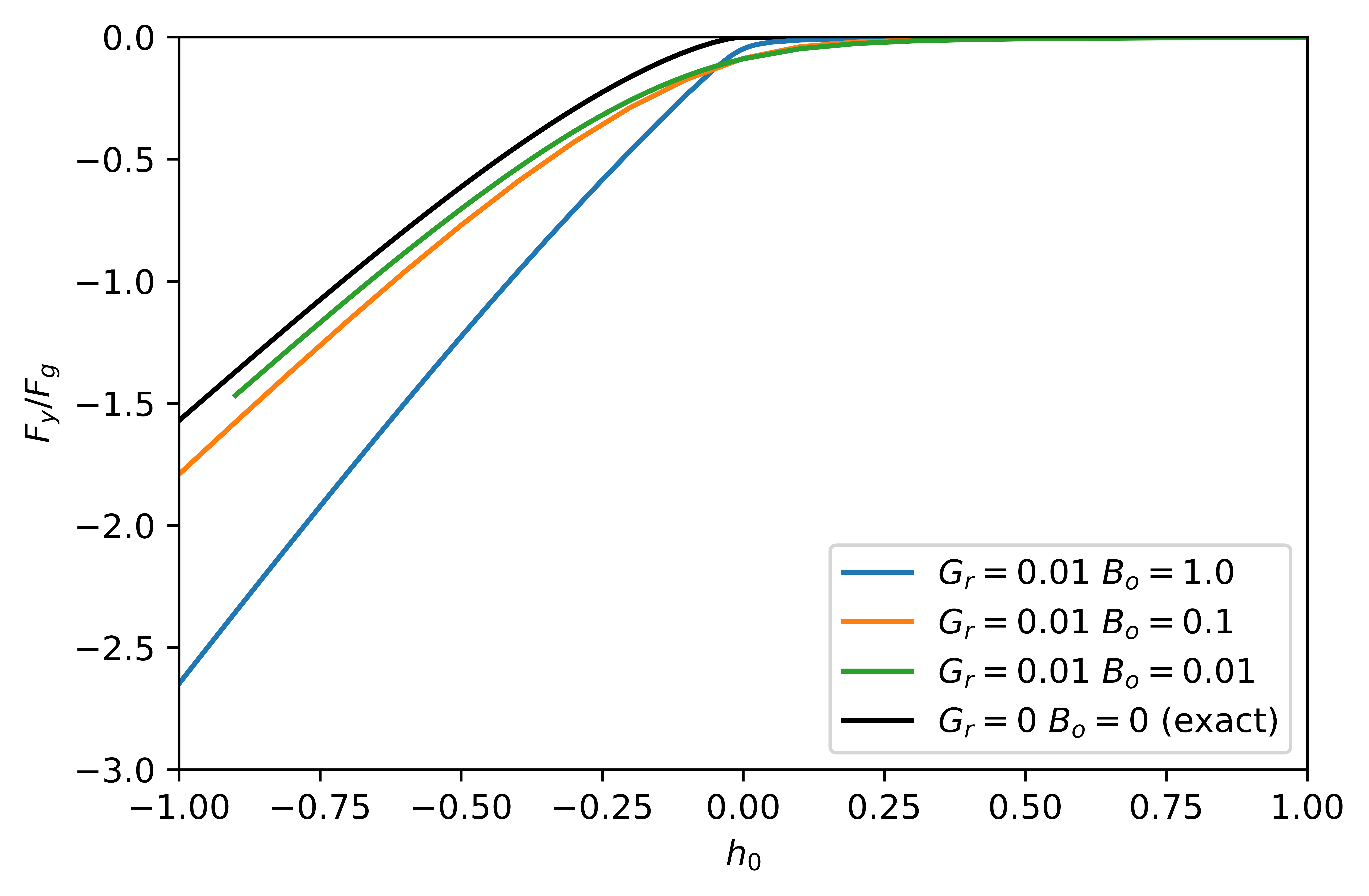}
        \includegraphics[width=0.45\columnwidth]{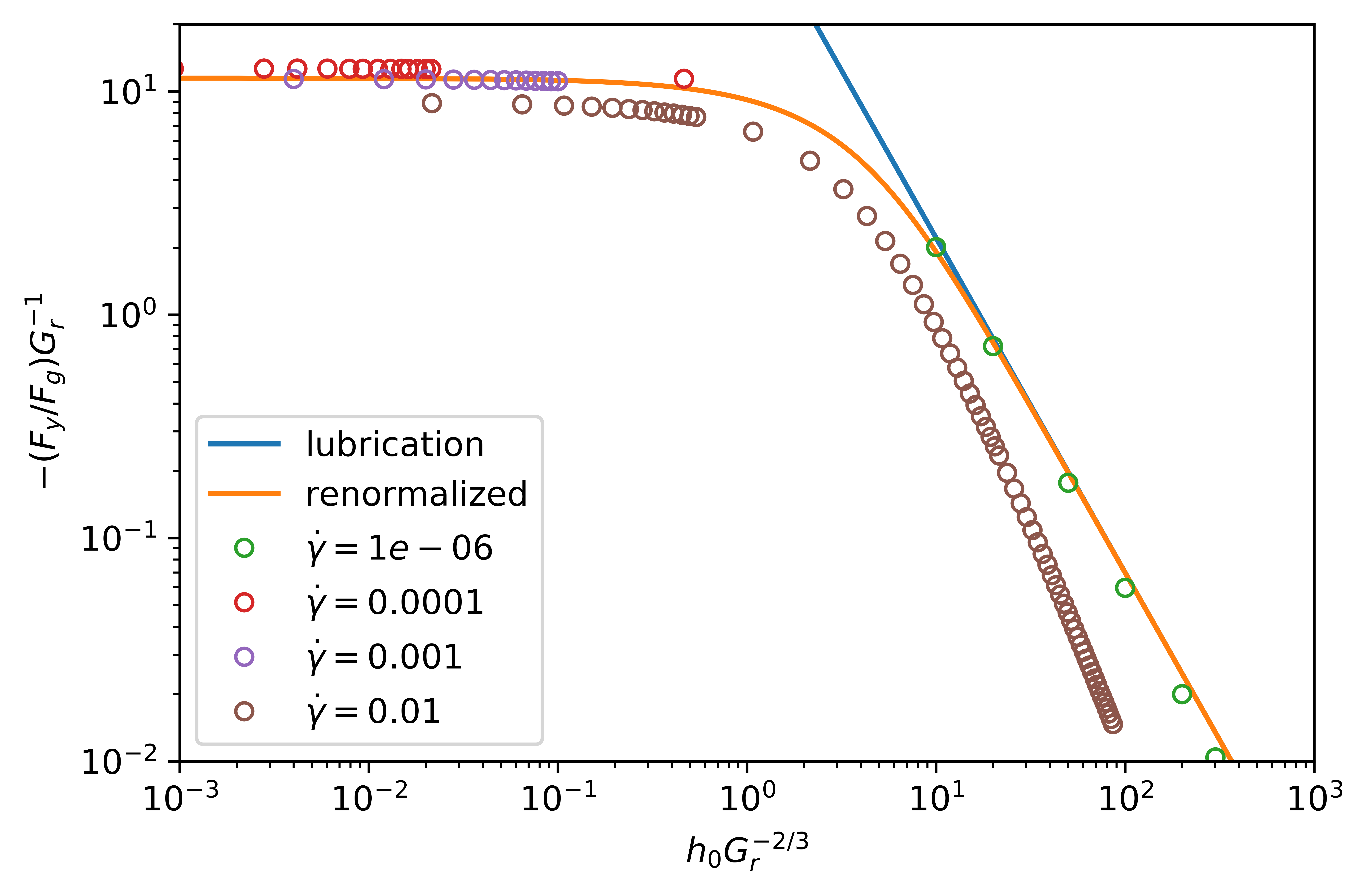}
        \caption{\label{fig:indent_Bond} Results for the $\sigma\rightarrow 0^+$ limit. Left: Indentation curves for $\sigma\rightarrow 0^+$. $F_g=\Delta\rho g a^2$ is the gravity force unit, $C_g=\dot\gamma\eta/(\Delta\rho g a)$ is the non-dimensional shear rate for the gravity-based interface elasticity Right:.
        }
    \end{center}
\end{figure}

Figure \ref{fig:indent_Bond} summarizes the results obtained for the zero-tension limit.
In the absence of surface tension, the downward force on the particle is non-dimensionalized by a characteristic gravity force $F_g=\Delta\rho g a^2$ and the shear rate is non-dimensionalized as the gravity number $G_r=\dot\gamma\eta/(\Delta\rho ga)$.

As given by eqs. (\ref{Taylorh2D}) and (\ref{TaylorF2D}), the indentation force scales as $F_y/F_g\approx -2(-2h_0)^{3/2}/3$ for $h_0<0$ and $\dot\gamma=0$ and $\sigma=0$.
This is reflected in the indentation curves under flow shown in Fig.\ref{fig:indent_Bond}, left panel.
Furthermore, the scaling of the lift force with $h_0$ is also qualitatively changed by setting $\sigma$ to 0, as given by eq. (\ref{classic_Fourier_gravity}).
As a consequence, the scaling exponents near the $h_0=0$, $G_r=0$ point are different from the $\sigma>0$ case:
\begin{equation}
    \label{scale_gravity}
    F_y/F_g=G_r\mathcal F_g(h_0/G_r^{2/3}),
\end{equation}
where $\mathcal F_g(\xi)\propto \xi^{-3/2}$ for $\xi\rightarrow\infty$ and $\mathcal F_g(\xi)\propto \xi^{3/2}$ for $\xi-\rightarrow\infty$.

The renormalization approach yields the following expression for $\sigma=0$
\begin{equation}
    \label{renom_gravity}
    F_y/F_g=-\frac{63\pi G_r^2}{\left[2h_0+|3F_y/(2F_g)|^{2/3}/2\right]^{3/2}},
\end{equation}
which, unlike the case of $\sigma>0$, shows only a qualitative agreement with the numerical results:
While the scaling exponents are correct for $h_0$ close to 0, the scaling constant at $h_0=0$ is underestimated by less than 20\% in Fig.\ref{fig:indent_Bond}, right panel.
This discrepancy is expected since the indented interface is far from being flat in the lubrication region if $\sigma=0$.

\end{document}